\documentclass[11pt]{article}
\usepackage{graphicx} 
\usepackage{amsfonts}
\usepackage{hyperref}
\usepackage{authblk}

\title{Drivers of Variation in the Optimal Spatial Structure of Collective Information Gatherers}
\author[1,2,*]{Ross S. Walker}
\author[3,4]{Gabriel Ramos-Fernandez}
\author[5]{Denis Boyer}
\author[3]{Sandra E. Smith-Aguilar}
\author[1]{Xander O'Neill}
\author[2]{Matthew J. Silk}

\affil[1]{Department of Mathematics and Maxwell Institute for Mathematical Sciences, Heriot-Watt University, Edinburgh, EH14 4AS United Kingdom}
\affil[2]{Institute of Ecology and Evolution, University of Edinburgh, Edinburgh, EH9 3FL United Kingdom}
\affil[3]{Instituto de Investigaciones en Matemáticas Aplicadas y en Sistemas, Universidad Nacional Autónoma de México, Mexico City, 04510 Mexico}
\affil[4]{Global Research Centre for Diverse Intelligences, University of St. Andrews, St Andrews, KY16 9AJ United Kingdom}
\affil[5]{Instituto de Física, Universidad Nacional Autónoma de México, Mexico City, 04510 Mexico}
\affil[*]{Corresponding author: r.s.walk02@gmail.com}

\usepackage[a4paper, total={6.5in, 9in}]{geometry}
\usepackage{mathtools}
\usepackage{enumitem} 
\usepackage{amsmath}
\usepackage{amssymb}
\usepackage{amsthm}
\usepackage[sorting = none, maxnames = 2, backend=biber, uniquename=false, maxbibnames=6, minbibnames=6, doi=true, url=false, giveninits=true, style=numeric-comp]{biblatex}

\DeclareSourcemap{
  \maps[datatype=bibtex]{
    \map{
      \step[fieldset=issn, null]
      \step[fieldset=isbn, null]
      \step[fieldset=month, null]
      \step[fieldset=language, null]
      \step[fieldset=number, null]
    }
  }
}

\date{}

\DefineBibliographyStrings{english}{%
  page             = {},
  pages            = {},
} 

\renewbibmacro{in:}{}
\hypersetup{
	colorlinks=true,
	linkcolor=black,
	citecolor=blue,
	urlcolor=blue}
\usepackage{geometry}

\begin{document}
\maketitle

\begin{abstract}
Collective systems that self-organise to maximise the group's ability to collect and distribute information can be successful in environments with high spatial and temporal variation. Such organisations are abundant in nature, as sharing information is a key benefit of many biological collective systems, and have been influential in the design of many artificial collectives such as swarm robotics. Understanding how these systems may be spatially distributed to optimise their collective potential is therefore of importance in both ecology and in collective systems design. Here, we develop a mathematical model which uses an optimisation framework to determine the higher-order spatial structure of a collective that optimises group-level knowledge transfer. The domain of the objective function is a set of weighted simplicial sets, which can fully represent the spatial structure from a topological perspective. By varying the parameters within the objective function and the constraints, we determine how the optimal spatial structure may vary when individuals differ in their information gathering ability and how this variation differs in the context of resource constraints. Our key findings are that the amount of resources in the environment can lead to specific subgroup sizes being optimal for the group as a whole when individuals are homogeneous in their information gathering abilities. Further, when there is  variation in information gathering abilities, our model implies that the sharing of space between smaller subgroups of the population, rather than the whole population, is optimal for collective knowledge sharing. Our results have applications across diverse contexts from behavioural ecology to bio-inspired collective systems design.\\


\end{abstract}

\section{Introduction}

Diverse natural and artificial systems are organised to optimise the transfer and distribution of information across a group of individuals. Collectively intelligent systems such as robot swarms, characterised by a lack of central control, simple movement rules and local interaction patterns, are increasingly used across multiple complex applications \cite{Schranz2020}, such as environmental monitoring \cite{Duarte2016}, targeted drug delivery \cite{Banharnsakun2016} and space exploration \cite{Escoubet2001}. Many of these artificial systems, designed for collection and distribution of information in complex, heterogeneous environments, are directly inspired by animal systems (such as eusocial insects; \cite{Schranz2020, Banharnsakun2016, Krieger2000}). 

In these animal systems, information sharing is one of the key drivers of group-living \cite{Evans2016, Brown1990}. While group-living comes with a variety of associated costs, such as increased competition for resources \cite{Alexander1974} and risk of pathogen transmission \cite{Romano2020, Sah2018}, these costs are mitigated through the many benefits of group-living \cite{Uetz2002, Salguero-Gómez2024, Ebensperger2016}, one of which is the sharing of foraging information \cite{Brown1986, Ebensperger2016a}. The sharing of foraging knowledge among conspecifics is an especially significant benefit of group-living for species living in highly heterogeneous environments with a high degree of spatial and temporal variation in the quality of feeding sites \parencite{Ebensperger2016a, Bhattacharya2014, Dunbar1988, Clark1984, Kummer1971, Moussaid2009} (e.g. tropical forests for frugivorous species \parencite{Smith-Aguilar2016}). In such systems, the pooling of knowledge between conspecifics facilitates a more complete tracking of the current foraging environment than what individuals could manage on their own  \cite{Moussaid2009, Bonabeau1999, Papageorgiou2024}. 

Information sharing between individuals in a collective system can take a variety of forms. In swarm robotics, for example, individuals transmit their current knowledge through short-distance signals, requiring spatial proximity \cite{Schranz2020}. In animal systems, information sharing can take a wide variety of forms, such as vocalisations (e.g. in meerkats (\textit{Suricata suricatta}); \cite{Gall2017}), observation (e.g. in guppies (\textit{Poecilia reticulata}) \cite{Laland1997}) or following (e.g. in hooded crows (\textit{Poecilia reticulata}); \cite{Sonerud2001}), and can be either active/intentional or passive/involuntary. In each of these cases, information transfer could potentially be costly to individuals in the short-term (due to resource sharing), but this cost can be tolerated for future reciprocity \cite{Wilkinson1984, Carter2013} (depending upon total resource availability, which may determine the effective benefit of less cooperative strategies; we may expect weaker kin selection in systems with high within-group competition for resources). Spatial proximity is typically required for information sharing in many animal species, which suggests that the spatial distribution of the collective is of intrinsic importance in the efficiency of such sharing. Therefore, some spatial structures may promote more effective information processing than others in collective systems. Given that there is a relationship between space sharing and information processing, it is reasonable to suggest that different spatial structures may promote more effective information processing in collective systems. Key pre-existing ecological hypotheses relating spatial structure and information processing are the \textit{recruitment centre hypothesis} \cite{Evans2016} and \textit{information centre hypothesis} \cite{Evans2016, Richner1996}, which both propose that spatial congregations, typically including all group members in a shared central area, allow for effective collective information processing, in turn allowing the group to be more adaptive to ecological and environmental variation. However, to the best of our knowledge, it is currently an open question which spatial structures are \textit{optimal} for collective information processing. 

One important aspect of natural collectives is within-group heterogeneity in group foraging abilities, which we may expect particularly in an ecological context (although see \cite{Kaminka2025}). In animal systems, there may be variation in individual abilities to move \cite{Nathan2008}, or willingness to gather or provide information \cite{Jeanson2013}, which translate to variation in abilities to forage or gather other kinds of information. Many factors may influence this variation, such as differences in metabolic rate \cite{Biro2010}, cognitive ability \cite{Manattini2024, Kashetsky2021}, social position \cite{Marshall2015, Stephens2007, Marshall2012}, social affinities \cite{Nathan2008}, personality \cite{Aliperti2021}, age \cite{Martins2024, Froy2015} or sex \cite{Reyes-González2021, Ruckstuhl2006}. If the group is to organise to maximise their collective potential for information sharing, then these heterogeneities should be accounted for. Such an optimal organisation might be expected when the individual benefits of information sharing outweigh the costs \cite{Parker1986, Robinson1986, Romano2022}, with the importance of this (specifically collective) optimisation enhanced when multilevel selection acts on group-level benefits and we expect kin selection \cite{Fisher2017, Philson2024}. For example, under the optimal strategy for the group as a whole, a forager with greater foraging abilities than the rest of the group may share more information than others, possibly leading to the costs of this behaviour outweighing the benefits for said individual. In such a scenario, we may not expect the (group-level) optimal structure to evolve due to unequal distribution of individual-level costs, and the forager with greater foraging abilities may even seek out a different group, or opt for solitary living \cite{Brown1990, Robinson1986}. It is therefore important to examine how variation in foraging ability may translate to differences in both the spatial composition which is optimal for information sharing at the group-level, and in the individual-level costs to this spatial organisation. 

Another important aspect of spatial composition is \textit{group size}, or the number of individuals within the collective system. In robot swarms, the number of robots is determined by the type and size of the task and by project budgets, with the costs of single robots varying dramatically between applications  \cite{Schroeder2019}. Theoretical studies suggest that there is an optimal swarm quantity because, while increasing the number of robots initially improves task performance, there can be diminishing marginal returns upon increase of group size due to possible collisions and task interference \cite{Lerman2002}. In animal systems, there is considerable variation in group size both between and within species \cite{Brown1990, Chapman1995}, and this variation is a key aspect in the balance of the costs and benefits to group-living and social behaviour as a whole \cite{Silk2007, Brown1990, Markham2015, Chapman2015, Wittenberger1985}. In particular, larger groups in resource-constrained environments may be more prone to within-group competition \cite{Lee2018, Seiler2020, Alexander1974} and be associated to reduced individual reproductive rates \cite{Borries2008}, but may be more effective at collecting and sharing foraging information \cite{Gibbs1987, Cantor2020, Giraldeau1984, King2007}. Indeed empirical studies have found a positive relationship between resource availability and group size \cite{Seiler2020, Gibbs1987}, which may be related to the decreased influence of resource competition in more abundant environments. In sum, group-size is a vital component in a collective's capacity to process and transfer information, and larger groups may not always be more efficient. It is therefore important to consider both how it may influence the properties of the optimal group spatial structure and how operating under optimal regimes may select for a particular group size.

In \cite{Ramos-Fernandez2025}, we propose that the spatial structure of the group that is optimal for sharing knowledge may represent a balance between the collective benefit of further exploration of individuals and their opportunities for sharing with others. Specifically, this balance implies that the intersection between a set of individual areas should be large enough for them to coincide and exchange knowledge but not so large that they do not have any unique areas known to share information about. As a component of this model we quantified the amount of knowledge sharing for a certain class of spatial structures and used an optimisation approach to estimate the optimal structure (to be used as a baseline for comparison with the empirically observed spatial structure of a group of Geoffroy's spider monkeys (\textit{Ateles geoffroyi})). However, this optimality framework assumed that all individuals were equal in their foraging ability and that there was no constraint on resource availability, despite the importance of these factors in real systems. Considering these additional factors, and how they influence spatial structures under an optimal regime, could improve our general knowledge regarding the drivers of variation in natural spatial structures of foraging groups, for which we lack general theory.

In this study we generalise and formalise the optimisation component within the model of \cite{Ramos-Fernandez2025} to determine the spatial structure of a group of independent foragers which optimises the group-level transfer of knowledge when individuals vary in their information gathering ability and are placed in resource-constrained environments. The optimal structure is determined as the optimal balance between the opportunities for individuals to coincide and the relative uniqueness of their knowledge, represented here as the solution to a constrained integer-valued optimisation problem. Our model utilises an abstraction of the information environment into an arbitrary collection of foraging sites, allowing us to draw more broad (topological) conclusions which are independent of any specific geometry. Each variable quantifies the number of `resource' points shared uniquely within the core area overlap of any possible subset of individuals. By exploring the parameter space of this model we consider: 1) the impact of variation in foraging abilities upon the optimal spatial composition of the group; 2) the impact of resource sparsity upon this structure; and 3) how these results vary with group-size, whenever computationally feasible. This paper is structured as follows: in Section \ref{sec: methods} we describe our model, the optimisation procedure and our methods for exploring parameter space. In Sections \ref{sec: homo}, \ref{sec: unique} and \ref{sec: hetero} we explore model outcomes in three `scenarios', each varying model parameters in a different way. In Section \ref{sec: conclusions} we discuss our key model outcomes in the context of behavioural and evolutionary ecology, highlighting potential applications towards more general systems of collective intelligence. In an ecological context, our model reveals the significant role of individual foraging abilities in the spatial structure of groups of all sizes when organising solely for optimal knowledge transfer, and shows how the importance of differences in foraging abilities covaries with resource availability foraging environments. While our model, as with many artificial collective systems, is inspired by animal social behaviour it has broad applications beyond ecology, for example in the design of swarm robotic systems.

\section{Methods}\label{sec: methods}

We represent the higher-order spatial structure of a collective system which maximises their group-level information transfer as the solution to a constrained mixed integer quadratic programming (MIQP) problem. Particularly, the optimal spatial structure is the maximiser of some information transfer function, $T$, subject to some constraints reflecting resource abundance and individual capacities for sharing. This solution describes how knowledge of foraging sites is distributed between subsets of individuals without reference to the spatial distribution of the points of knowledge themselves. The solution therefore provides a \textit{topological} description of the optimal group spatial structure. We develop quantitative measures of the optimal spatial structure, in order to conceptualise and contrast different higher-order structures. These allow us to study how individual abilities to forage, represented through knowledge of foraging points, and the amount of resources of the environment may drive variation in optimal spatial structures. We consider this variation through three `scenarios', with sequentially increasing within-group heterogeneity in foraging ability.

\subsection{Problem variables and parameters}

Let $n$ be the number of individuals within the environment/group. The foraging environment is represented as a collection of $F\in\mathbb{N}$ many points (taking the convention $0\notin \mathbb{N}$), each representing a potential foraging site. Each individual $i$ has a knowledge capacity of $N_i\in \mathbb{N}$ many points for $i\in [n] \vcentcolon = \{1,\dots,n\}$. We assume that individuals utilise their knowledge capacity as far as their environment will allow, so an individual with capacity $N_i$ will have \textit{effective} knowledge of $N_i'\in\mathbb{N}$ many points, where
\begin{equation*}
    N_i' = \min\{N_i, F\}.
\end{equation*}
Each point can be known uniquely by either; 1) zero individuals, 2) a single individual, or 3) a \textit{subgroup} of individuals (here exclusively referring to groups of at least 2 individuals). 

\begin{figure}[t!]
    \centering
    \includegraphics[width=\linewidth]{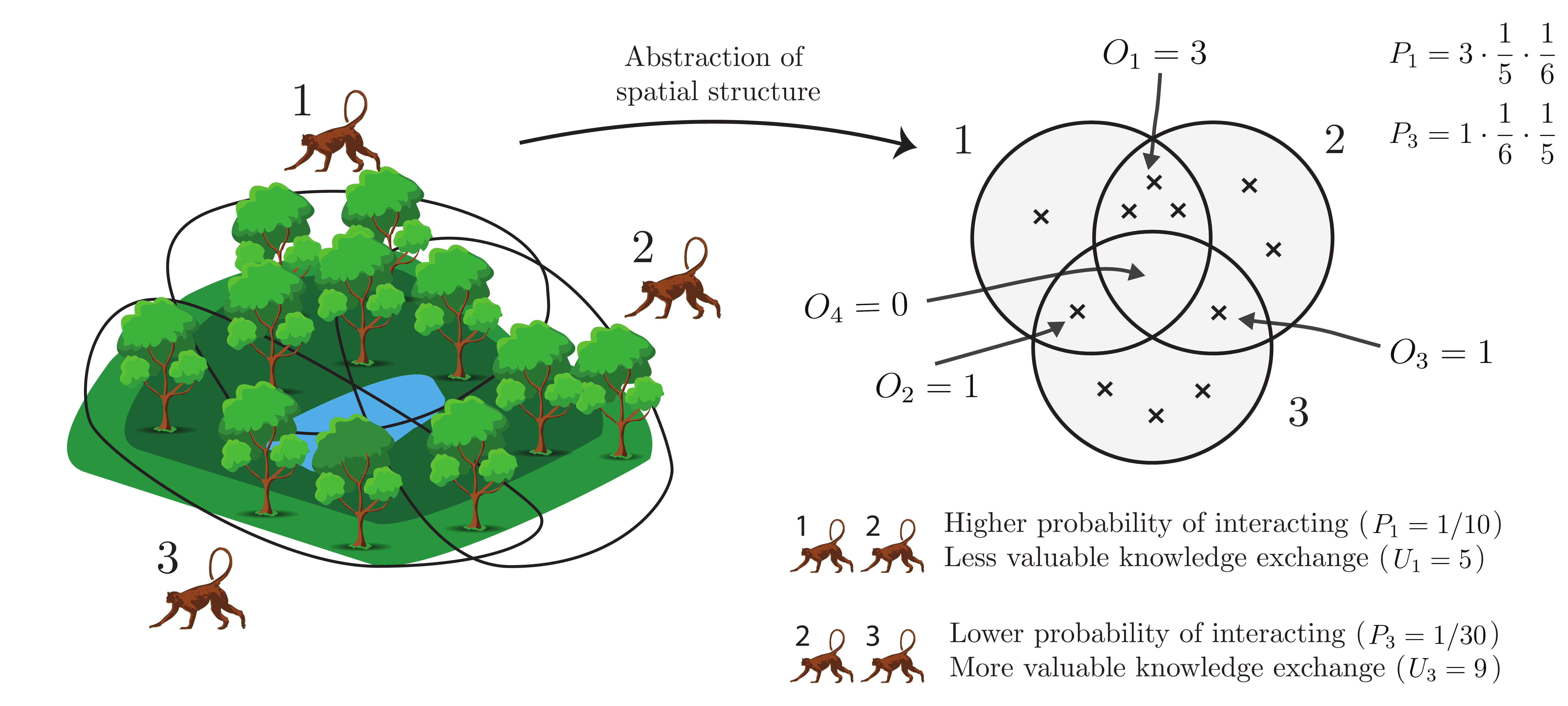}
    \caption{Conceptualisation of the foraging environment and problem variables. The knowledge points represent, in this case, the trees in which food may be available. In this example, individuals 1 and 2 have relatively less unique knowledge, but higher probability of interaction, in contrast to individuals 2 and 3.}
    \label{fig: methods_fig}
\end{figure}

The set of all possible subgroups, $\mathcal{X}$, is obtained as the power set of $[n]$ minus singletons and the empty set, taking cardinality $|\mathcal{X}| = \mathcal{N}(n) = 2^n-n-1$, which is the dimensionality of our optimisation problem. We order this set by first arranging the elements in ascending cardinality (so interactions between fewer individuals come \textit{before} interactions between more individuals) and then arranging elements of the same cardinality lexicographically. Denote the bijection corresponding to this ordering by $I\colon \mathcal{X}\to [\mathcal{N}]$. The variables of the optimisation problem are $\boldsymbol{O}=(O_i)\in\mathbb{N}^{\mathcal{N}}$, where $O_i$ gives the number of points uniquely known by the subgroup with index $i$ under enumeration $I$. These variables are exemplified in Figure \ref{fig: methods_fig}.
Note that $\boldsymbol{O}$ fully describes the higher-order spatial structure, since the number of cells known by only a single individual $i$ can then be determined by their own effective foraging knowledge, $N_i'$ minus their number of shared points within subgroups, and then the number known by zero individuals can then similarly be determined from $F$.

\subsection{Objective function}\label{sec: objective}
We assume that individuals move between their points of knowledge uniformly, so the probability that the $i$-th individual occupies a particular point at a given time is
\begin{equation*}
    p_i = \frac{1}{N_i'}.
\end{equation*}
Then, assuming that individuals move independently, the probability of the $c$-th subgroup (under enumeration $I$) coinciding at the same foraging point under the spatial structure $\boldsymbol{O}$ is given as
\begin{equation*}
P_c\left(\boldsymbol{O}\right) = \left(\prod_{k\in I^{-1}(c)}p_k\right)\left( \sum_{a\in S(c)} O_a\right)  = \frac{\sum_{a\in S(c)} O_a}{\prod_ {k\in I^{-1}(c)}N_k'}
\end{equation*} 
where 
\begin{equation*}
    S(c) = \left\{i\in \left[\mathcal{N} \right] : I^{-1}(c)\subseteq I^{-1}(i)\right\}
\end{equation*}
gives the indices of subsets which include the $c$-th subgroup for all $c\in\left[\mathcal{N}\right]$. Under this construction, any subgroup of individuals can interact in both the areas that they uniquely have knowledge of \textit{and} in the areas known by them and additional individuals. We then assume that when a subgroup meets they share their \textit{collective information}. This is defined as the number of points known by at least one individual in the subgroup, but not by all individuals. For the $c$-th subgroup, this is given by
\begin{equation*}
    U_c(\boldsymbol{O}) = 
    \underbrace{\sum_{k\in I^{-1}(c)}N_k'}_{\text{Total knowledge in }I^{-1}(c)} - \underbrace{f(c)\sum_{a\in S(c)}O_a}_{\text{Points known by \textit{all} of  }I^{-1}(c)}-\underbrace{\sum_{a\in B(c)}(f(a)-1)O_a}_{\text{Over-counted points}}
\end{equation*}
where
\begin{equation*}
B(c) = \left\{i\in \left[\mathcal{N} \right] : I^{-1}(i) \subset I^{-1}(c)\right\}
\end{equation*}
is the set of indices of all the smaller subgroups of individuals contained in the $c$-th subgroup of individuals under the enumeration $I$, and $f\colon [\mathcal{N}]\to [n]$ is the size of the $c$-th subgroup, $f(c)=|I^{-1}(c)|$. 

We then assume that the rate of information transfer when the subgroup coincides is directly proportional to this uniquely known area. The net information transfer across the entire group of $n$ individuals is then given by $T\colon \mathbb{R}^{\mathcal{N}(n)} \to \mathbb{R}$
\begin{align*}
    T(\boldsymbol{O}) &= \sum_{c=1}^\mathcal{N} P_c(\boldsymbol{O})U_c({\boldsymbol{O}}) \\
    & = \sum_{c=1}^\mathcal{N} \left(\frac{\sum_{a\in S(c)} O_a}{\prod_ {k\in I^{-1}(c)}N_k'}\right)\left(
\sum_{k\in I^{-1}(c)}N_k' - f(c)\sum_{a\in S(c)}O_a-\sum_{a\in B(c)}(f(c)-1)O_a
    \right).
\end{align*}
Observing that this is a quadratic form, we can write
\begin{equation*}
    T(\boldsymbol{O}) = \boldsymbol{l}^t\boldsymbol{O}+\frac{1}{2}\boldsymbol{O}^tM\boldsymbol{O} 
\end{equation*}
for some $\boldsymbol{l}\in \mathbb{R}^{\mathcal{N}}$ and $M\in\mathbb{R}^{\mathcal{N}\times\mathcal{N}}$. In Section S1.1 we explicitly derive the coefficients of the vector $\boldsymbol{l}$ and matrix $M$.

\subsection{Constraints}\label{sec: constraints}

Our problem constraints are determined by the following; 1) Individuals must share a non-negative number of points within subgroups, 2) individuals cannot share more points than they have knowledge of, and 3) the group as a whole cannot have knowledge of more than $F$-many foraging points. Respectively, this yields the constraints
\begin{align}
    &\boldsymbol{O}\geq \boldsymbol{0}, \nonumber\\
    & \sum_{i\in \tilde{S}(k)}O_i\leq N_k', \text{ for all } k\in [n], \label{eq: sharing constraint}\\ 
    & \sum_{i=1}^nN_i' - \sum_{c=1}^\mathcal{N}(f(c)-1)O_c  \leq F\label{eq: forage constraint},
\end{align}
where $\tilde{S}(k)$ is the set of indices of subgroups which contain the $k$-th individual, given explicitly by
\begin{equation*}
    \tilde{S}(k) = \left\{i\in[\mathcal{N}]: \{k\}\subset I^{-1}(i)\right\}.
\end{equation*}
Equation \eqref{eq: forage constraint} sums over each individual's effective knowledge and then accounts for over-counted points due to knowledge sharing under the spatial structure $\boldsymbol{O}$.

Whenever $F\geq\sum_{i=1}^nN_i'$ the constraint \eqref{eq: forage constraint} is automatically satisfied by the non-negativity of $\boldsymbol{O}$. We refer to this case as `resource-abundant', and the case where $F<\sum_{i=1}^nN_i'$ as `resource-constrained'. The constraints \eqref{eq: sharing constraint} and \eqref{eq: forage constraint} are both linear, so can be collectively written in the form $G\boldsymbol{O} \leq h$, where $G\in \mathbb{R}^{(n+1)\times \mathcal{N}}$ and $h\in\mathbb{R}^{n+1}$. The specific entries of $G$ and $h$ are given in Section S1.2. 

\subsection{Optimisation procedure}
The maximisation problem of the objective function, $T$, defined in Subsection \ref{sec: objective} subject to the constraints defined in Subsection \ref{sec: objective}, written as
\begin{align}\label{eq: main problem}
\max  _{\boldsymbol{O}}             &\; \quad T(\boldsymbol{O}) = \boldsymbol{l}^t\boldsymbol{O}+\frac{1}{2}\boldsymbol{O}^tM\boldsymbol{O} \\
\text{s.t} &\;
\begin{alignedat}[t]{1}
                            \quad G\boldsymbol{O}& \leq h \\
                             \boldsymbol{O} & \geq \boldsymbol{0} \\
\boldsymbol{O}&\in\mathbb{Z}^{\mathcal{N}(n)}                    
\end{alignedat}
\end{align}
is feasible whenever $F\geq0$ (see Section S1.3). We solve this MIQP using a suite of global optimisation solvers: \textit{Gurobi} \cite{Gurobi}, \textit{IBM ILOG CPLEX}, and \textit{FICO Xpress}, all obtained through their free academic licences and implemented in Python. Each of these solvers utilise a kind of branch-and-cut method for solving MIQPs. With each combination of parameters (i.e. each value of $F$, $n$ and $\boldsymbol{N}$) considered in this study, the corresponding optimisation problem is solved independently with each solver, and the `best' solution (that which provides the largest maximal value of $T$), $\boldsymbol{O}^*$, is returned. 

\subsection{Measures of the spatial structure}
We develop descriptive measures of the higher-order spatial structure. For $i=2,\dots, n$ we define the \textit{relative $i$-th order overlap} of the spatial structure as
\begin{equation*}
    w_n^{(i)}(\boldsymbol{O}) = \frac{\sum_{c=1}^\mathcal{N}I_i(f(c))O_c}{ \sum_{i=1}^nN_i' - \sum_{c=1}^\mathcal{N}(f(c)-1)O_c}
\end{equation*}
where $I_i\colon [n]\to \{0, 1\}$ is the indicator function (with $I_i(x)=1$ if $x=i$ and $I_i(x)=0$ otherwise). Therefore, $w_n^{(i)}$ describes the proportion of foraging points known uniquely by subgroups of size $i$ (relative to the number of points known by at least one individual in the group). The proportion of points which are known by the \textit{entire} group is then $w_n^{(n)}$. The proportion of points known by \textit{exactly one} individual, $w^{(1)}_n$, can then be obtained by the remaining proportion of foraging points
\begin{equation*}
    w^{(1)}_n(\boldsymbol{O}) = 1 - \sum_{i=2}^nw_n^{(i)}(\boldsymbol{O}).
\end{equation*}
We can express the \textit{expected number of individuals with knowledge of each point}, $K(\boldsymbol{O})$, in terms of the proportions $w_n^{(i)}$ as
\begin{equation}
    K(\boldsymbol{O}) = \sum_{i=1}^n i w_n^{(i)}(\boldsymbol{O}).
\end{equation}
Lower values of $K$ indicate that each point is known by fewer individuals on average (i.e. there might be more crowding of individuals around foraging points), which may imply lower individual-level costs as individuals typically share their resources with fewer individuals and therefore those resources may be of higher value (particularly in the context of food sharing). This metric can take values between 1 (when no space is shared by any individuals, so $w_n^{(1)} = 1$) and $n$ (when all space is shared by all individuals, so $w_n^{(n)} = 1$). 

Similarly, we quantify the proportion of points that the $k$-th individual shares with any other individual as
\begin{equation*}
    C_k(\boldsymbol{O}) = \frac{\sum_{i\in \tilde{S}(k)}O_i}{N_k'}
\end{equation*}
for $k\in [n]$. Higher values correspond to a higher proportion of points shared with others and therefore a lower proportion of uniquely known points, which may represent some individual-level cost \cite{Lee2016}, now specific to the k-th individual, occurred with the the spatial structure $\boldsymbol{O}$.

\subsection{Scenarios}\label{sec: scenarios}
We consider how the solution to the MIQP \eqref{eq: main problem} varies with individual abilities, $\boldsymbol{N}=(N_i)_{i\in [n]}$, and resource abundance, $F$, through the following scenarios: 
\begin{enumerate}
    \item \textit{Homogeneous knowledge scenario}, Section \ref{sec: homo}; all individuals have the same foraging ability, so $N_i=N$ for some $N\in\mathbb{N}$. 
    \item \textit{Distinct forager scenario}, Section \ref{sec: unique}; all individuals have the same foraging ability, $N_i=N\in\mathbb{N}$, except for one individual with ability $N_1=DN$ for $D\in \mathbb{R}^{>0}$. The impact of variation in the parameter $D$ is considered.
    \item \textit{Heterogeneous knowledge scenario}, Section \ref{sec: hetero}; individual abilities, $N_i$, are drawn independently according to a normal distribution with fixed mean $N$ and variance $\sigma\geq 0$ truncated so that $N_i>0$, and then rounded so that $N_i\in\mathbb{N}$ for each $i\in [n]$.
    The impact of variation in the parameter $\sigma$ upon the optimal spatial structure is considered through Monte Carlo simulation with 250 simulations for each value of $\sigma$.
\end{enumerate}

For each of these scenarios, we consider the role of both group size $n$ (where computationally feasible due to exponential growth in problem dimensionality with $n$) and resource abundance $F$ upon the optimal spatial structure, differentiating between the resource-abundant and resource-constrained scenarios.

\section{Homogeneous knowledge scenario}\label{sec: homo}
This scenario represents the simplest group composition, where all individuals have exactly the same ability to forage ($N_i=N\in\mathbb{N}$ for all $i\in[n]$). We discuss the resource-abundant case as a baseline for comparison with the resource-constrained case.  

\subsection{Resource-abundant environment}
In this case, our model is equivalent to that of \cite{Ramos-Fernandez2025}, with the additional formalisation of integer-valued variables and the inclusion of our additional descriptive measures. The solution to the information sharing optimisation problem \eqref{eq: main problem} is
\begin{equation}
    \boldsymbol{O}^* = \left(0,\dots, 0, \frac{N}{2}\right)
\end{equation} 
when $N$ is even, and a similar solution is returned when $N$ is odd, either rounding up or down the final entry of $N/2$\footnote{We consider the case where $N$ is even throughout this paper, emphasising that the results are almost identical in the odd case.}. In Section S2.1 we directly prove that $\boldsymbol{O}^*$ is the \textit{global} optima of $T$ in this homogeneous, resource-abundant case. This corresponds to an exact balance between exploration and sharing; each individual shares half of their space with \textit{all} other individuals in a centrally shared area, and the other half is not shared with any other individual, with no lower-order space sharing occurring. This aligns with the information centre hypothesis in ecology, which suggests that individuals share a common area as an effective means to collect and distribute information across the group \cite{Evans2016}. Here, the spatial structure described by $\boldsymbol{O}^*$ has the characteristics:
\begin{equation*}
    w_{n}^{(1)} = \frac{n}{n+1}, \hspace{10pt}w_n^{(n)} = \frac{1}{n+1}, \hspace{10pt} w_n^{(i)} = 0 \text{ for } i = 2, \dots, n-1,
\end{equation*}
\begin{equation*}
    K = \frac{2n}{n+1}, \hspace{10pt}C_k = \frac{1}{2} \text{ for all } k\in [n] 
\end{equation*}
with the corresponding information transfer
\begin{equation*}
    T(\boldsymbol{O}^*) = \frac{N^2}{2}\sum_{k=2}^n \binom{n}{k}
    kN^{-k} = \frac{n(n-1)}{4} + \mathcal{O}(N^{-1})
\end{equation*}
as derived in Section S2.2. Therefore, as the group size $n$ increases, the group knowledge transfer increases approximately quadratically (asymptotically with large $N$), while each individual maintains the same proportion of uniquely known points ($C_k$ is independent to $n$) and the group shares a lower proportion of their points ($w_n^{(n)}$ is decreasing in $n$). In larger groups, each point is known by more individuals on average ($K$ is increasing in $n$) due to the larger number of individuals in the shared central area. Although these changes in $w_n^{(n)}$ and $K$ become smaller in magnitude as $n$ increases due to the asymptotic nature of these quantities, and so the effect of increasing group size is diminishing.

\subsection{Resource-constrained environment}\label{sec:Homo_withF}

\begin{figure}[t!]
    \centering
    \includegraphics[width=\linewidth]{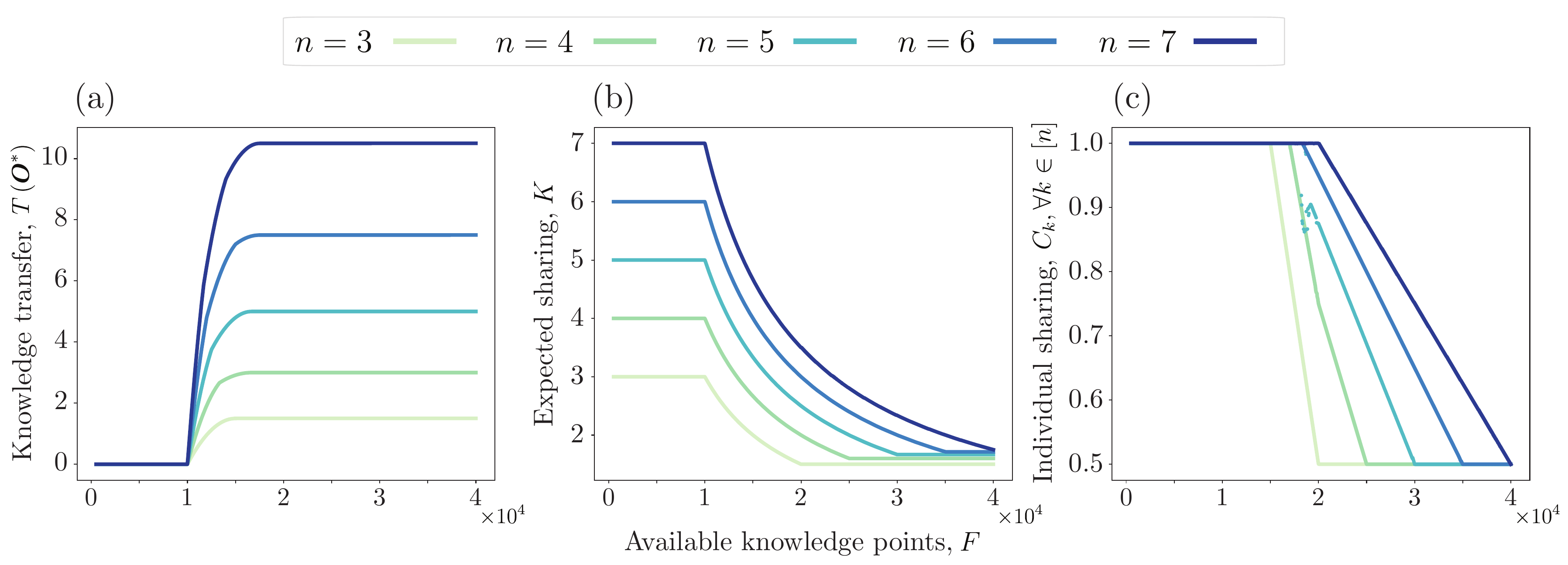}
    \caption{Quantitative descriptions of the optimal spatial structure in the homogeneous scenario ($N_i=N=10^4$ for all $i\in [n]$) with variation in resource abundance, $F\in [0.05\times 10^4, 4\times10^4]$. Specific values of $F$ are obtained from 1000 uniformly spaced points in this interval rounded upwards. Shown are: (a) the value of the objective function at the optimal solution, $T(\boldsymbol{O}^*)$, (b) the expected number of individuals with knowledge of each point under the optimal structure, $K(\boldsymbol{O}^*)$, and (c) the proportion of each individual's $N$-many points which are also known by other individuals, $C_k$ (which is fixed across $k\in [n]$ by problem symmetry). Darker colours correspond to higher group sizes, varying sequentially from $n=3$ to $n=7$.} 
    \label{fig: homo_noF_1}
\end{figure}

The optimal structure from the unconstrained foraging environment makes use of $N(n+1)/2$ many points ($N/2$ from the central shared space, and $nN/2$ from the remaining uniquely known points by all individuals), such that this solution remains feasible and optimal as long as $F\geq F^* =N(n+1)/2$. This critical value, $F^*$ is increasing with $n$, such that in larger groups the resource-abundant optimal structure remains feasible only with increasingly large values of $F$. 
As $F$ decreases below the critical value $F^*$, there is initially negligible increase in $T\left(\boldsymbol{O}^*\right)$ (Figure \ref{fig: homo_noF_1}a), such that groups are robust to decreases in the resource abundance of certain sizes. With this decreasing abundance there is an increase in $K$ (each point is known by more individuals due to the lower abundance of points) until $F=N$ (where $K$ reaches its maximum value of $K=n$; Figure \ref{fig: homo_noF_1}b). We also find that $C_k$ increases as $F$ decreases, so that in more harshly constrained environments individuals must share a higher proportion of their uniquely known points in order to maximise the group-level benefit (Figure \ref{fig: homo_noF_1}c). This additional sharing in the optimal structure appears rapidly with decreasing $F$, and individuals lose all of their unique knowledge ($C_k = 1$) before reaching the trivial structure observed at $F=N$ (compare Figures \ref{fig: homo_noF_1}b and \ref{fig: homo_noF_1}c). The change in the objective value $T$ when decreasing $F$ is small to begin with, even when $C_k=1$, but plummets to 0 when $F$ is close to $N$ and the completely-shared structure is forced (Figure \ref{fig: homo_noF_1}a). In this completely-shared structure, both $K$ and $C_k$ are maximal, but $T$ is minimal because no subgroup of individuals holds any unique information. This suggests that reorganisation to maintain group-level information transfer in adaptation to resource constraints is effective but that individual-level costs, such as decreased value of knowledge points, are incurred by forming this structure. All of these behaviours are qualitatively similar across tested group sizes (Figure \ref{fig: homo_noF_1}).

\begin{figure}[t!]
    \centering
    \includegraphics[width=\linewidth]{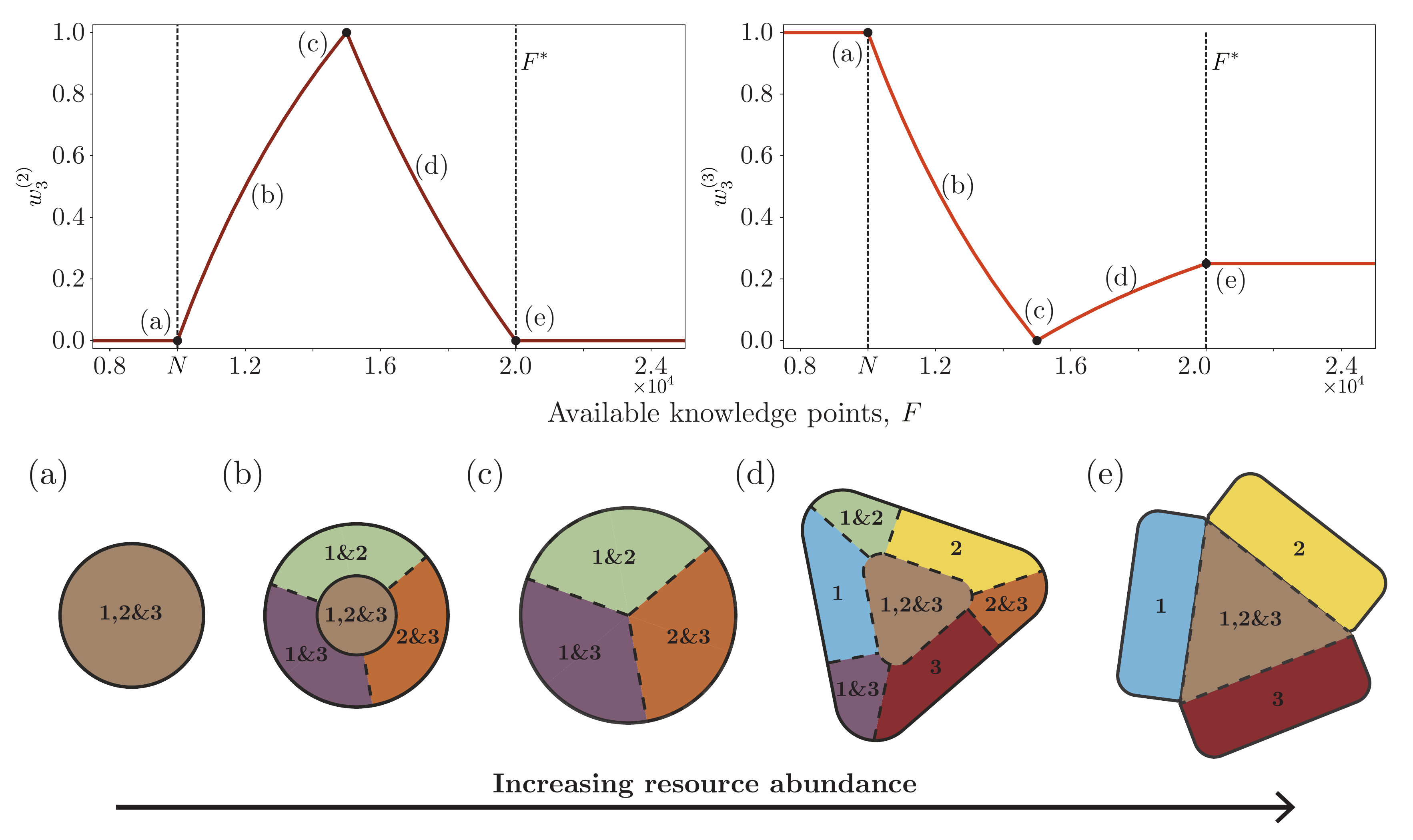}
    \caption{Variation in the proportion of higher-order overlaps with $F\in [7.5\times10^3, 2.5\times 10^4]$ in the homogeneous scenario ($N_i=N=10^4$ for all $i\in [n]$) for group size $n=3$. Specific values of $F$ are obtained from 1000 uniformly spaced points in this interval rounded upwards. Specific values of $F$ are obtained as 1000 uniformly spaced points in this interval, and rounding each value upwards). The critical value $F^*=N(n+1)/2$ for which the foraging constraint (Equation \eqref{eq: forage constraint}) is satisfied by the solution for the resource-abundant scenario for each $F\geq F^*$ is highlighted on the top plots. Shown are: (left) $w_3^{(2)}$, the proportion of dyadic point sharing and (right) $w_3^{(3)}$, the proportion of points uniquely shared between all three individuals. \textit{Conceptual} representations of the optimal spatial structures (using areas instead of points) corresponding to points along these plots are marked (a)-(e), with (e) representing the optimal spatial structure in the resource-abundant scenario, namely $\boldsymbol{O}^* = \left(0, 0, 0, N/2\right)$.}
    \label{fig:1b}
\end{figure}

\begin{figure}[t!]
    \centering
    \includegraphics[width=\linewidth]{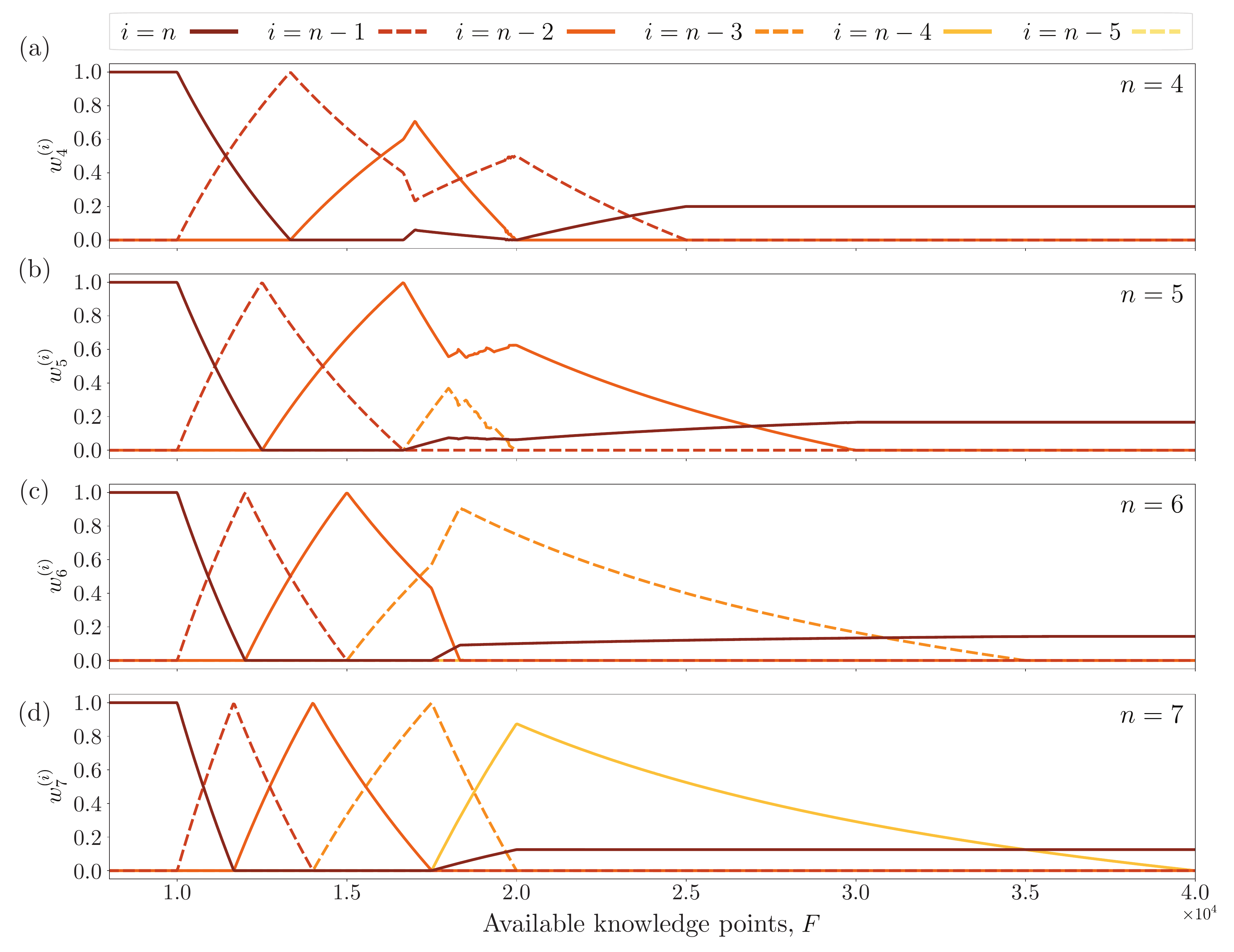}
    \caption{Variation in the order of spatial overlap, represented by $w_n^{(i)}$, in the homogeneous scenario ($N_i=N=10^4$ for all $i\in [n]$) with varying resource abundance $F\in[7.5\times 10^3, 4\times 10^4]$. Specific values of $F$ are obtained from 1000 uniformly spaced points in this interval rounded upwards. Group sizes are: (a) $n=4$, (b) $n=5$, (c) $n=6$, and (d) $n=7$. Darker colours correspond to orders of space sharing closer to $n$. The sequential dashed and solid lines are for plot clarity only.}
    \label{fig:homo_forage_ws}
\end{figure}

When initially decreasing $F$ from $F^*$ in the case of $n=3$, as visualised in Figure \ref{fig:1b}, there is initially a gradual switch between order-3 space sharing ($w_3^{(3)}$) and order-2 ($w_3^{(2)}$) space sharing, until the structure consists eventually entirely of dyadic space sharing ($w_3^{(2)}=1$). After this point, no lower-order interactions can be engaged in, so there is forced sharing between all group members which increases until a trivial structure is reached. This observation generalises to larger group sizes as follows. When initially decreasing $F$ from $F^*$ the proportion of maximal-order space sharing, $w_n^{(n)}$, decreases towards 0 and the proportion of some lower order spatial overlap, ($w_n^{(i)}$ for some $i<n$) increases towards some peak value (Figures \ref{fig:1b} and \ref{fig:homo_forage_ws}). When $F$ decreases beyond this peak in $w_n^{(i)}$, a similar switch occurs between the values of $w_n^{(i)}$ and $w_n^{(i+1)}$; the one-higher order spatial sharing takes place in the optimal structure instead of the lower-order. As the proportion of the lower-order space sharing reaches zero, the higher order reaches its peak, and the switch occurs again with a further higher-order sharing. This pattern continues until the peak in $w_n^{(n-1)}$, after which the proportion of centrally shared space $w_n^{(n)}$ increases from 0 to 1, indicating the forced sharing of space which occurs as $F$ approaches $N$. In short, particularly high or low values of $F$ can promote higher-order sharing, while intermediate values will typically promote lower-order sharing. We describe this as a \textit{pattern of interaction order selection}.

\section{Distinct forager scenario}\label{sec: unique}
In this scenario we considered variation in the foraging ability of just a single individual (called the \textit{distinct forager}), taking $\boldsymbol{N}=(DN, N, \dots, N)$ for some $D>0$. When $D<1$, the distinct forager has a lower ability (the \textit{one-worse} scenario), and when $D>1$ they have a higher ability (the \textit{one-better} scenario).  We consider the impact of variation in $D$ upon the optimal spatial structure of the group, and how this interacts with variation in resource availability $F$.

\subsection{Resource-abundant environment}\label{sec:Unique_noF}

\begin{figure}[t!]
    \centering
    \includegraphics[width=1\linewidth]{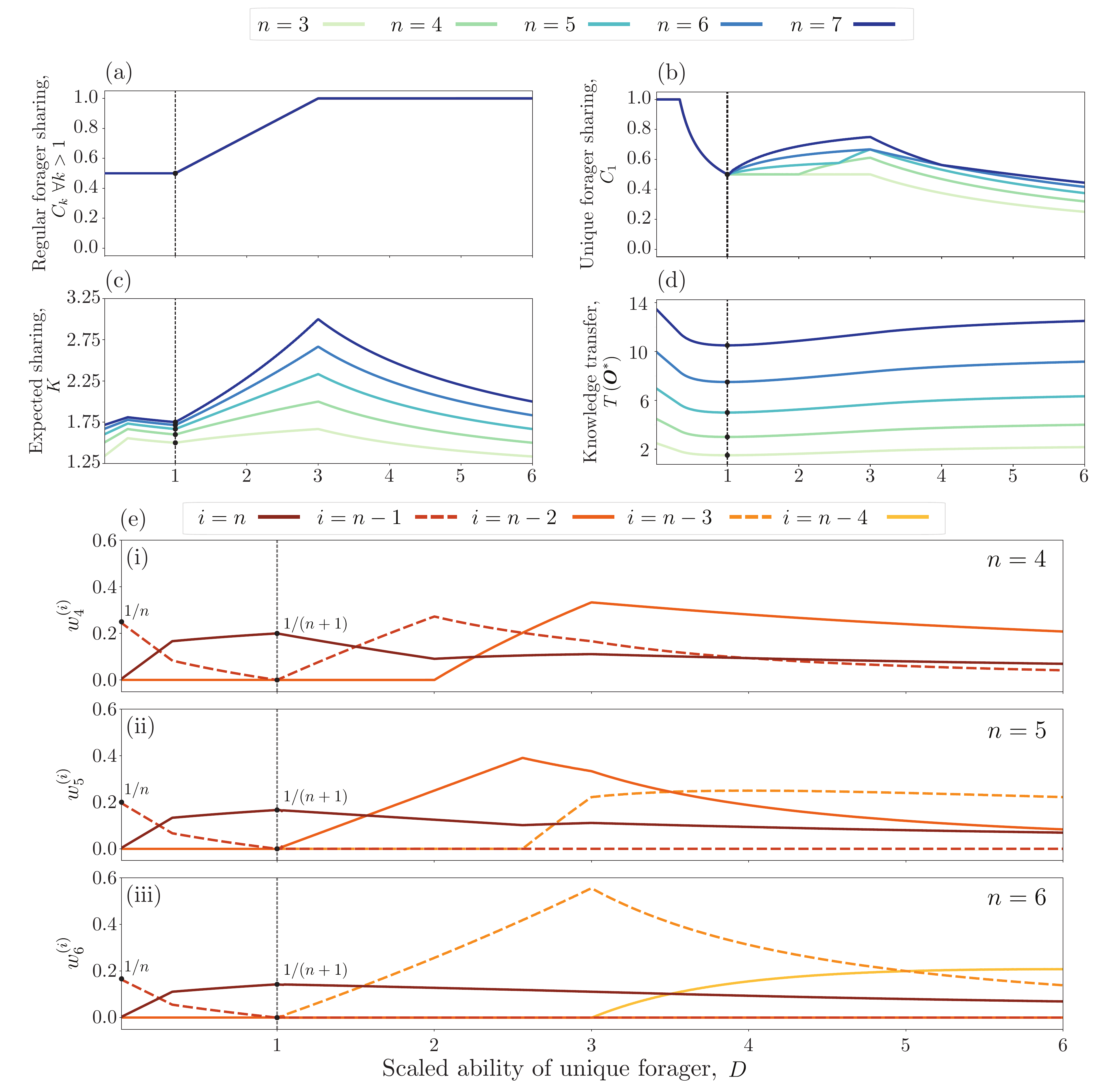}
    \caption{Results from the (resource-abundant) distinct forager scenario, where $\boldsymbol{N}=(DN, N, \dots, N)$ for $D\in [0.01, 6.00]$. Specific values of $D$ are obtained from 1000 uniformly spaced points in this interval rounded upwards. Shown are: (a) the proportion of regular individual's $N$-many points are also known by other individuals, $C_k$ (for $k>1$), (b) the proportion of the distinct foragers $DN$-many points also known by other individuals, $C_1$, (c) the expected number of individuals with knowledge of each point under the optimal structure, $K$, (d) the value of the objective function at the optimal solution, $T(\boldsymbol{O}^*)$, and (e) the relative orders of spatial overlap, $w_n^{(i)}$, for $i=2,\dots, n$ with group size (i) $n=4$, (ii) $n=5$, (iii) $n=6$. In (a)-(d), darker colours correspond to higher group size ($n=4,5,6$), while in (e) they correspond to orders of space sharing closer to $n$.}
    \label{fig:unique_no_F}
\end{figure}

In the one-worse scenario, as $D$ \textit{decreases} from $D=1$ (which represented the homogeneous scenario; Section \ref{sec: homo}), regular individuals do not sacrifice any more of their uniquely known points, represented by $C_k$ being constant for $k\geq 2$ and $D\in (0, 1]$, shown in Figure \ref{fig:unique_no_F}a. However, the distinct forager shares more of their own unique area and eventually all of it ($C_1$ is increasing until around $D=1/3$, at which point it takes its maximal value of $C_1=1$;  Figure \ref{fig:unique_no_F}b), suggesting that the optimal strategy (specifically at the group-level) is for the worse forager to be `supported' in an altruistic manner. However, in natural systems, since this behaviour is highly non-reciprocal, based on foraging behaviour alone we may not expect this kind of structure unless selection at the level of the group mostly drives spatial structure (rather than selection at the individual-level). Variation in the unique area held by each individual is exactly consistent between the tested group-sizes. As $D$ approaches $0$ in a group of size $n$, the optimal structure from a homogeneous group of size $n-1$ expectedly emerges. Particularly, $w_n^{(n-1)}$ approaches $1/n$ and all other values of $w_n^{(i)}$ (for $i>1$) approach 0 (see Figure \ref{fig:unique_no_F}e). 

In the one-better scenario, as $D$ \textit{increases} from $D=1$ (now representing the one-better scenario) both the distinct forager and the regular foragers share a higher proportion of their space with others (each $C_k$ is increasing; Figures \ref{fig:unique_no_F}a and \ref{fig:unique_no_F}b), leading to an increase in the expected sharing $K$ (Figure \ref{fig:unique_no_F}c). This indicates that the optimal strategy for the group is for the distinct forager to share more space with others rather than exploring more points which are novel to the group (relative to the homogeneous structure). The increased sharing continues until $D=3$, when regular individuals share all their space with the distinct forager ($C_k = 1$ for $k\geq 2$), maximising their probability of interaction.  Ecologically, this increased sharing implies that distinct foragers may end up in more central socio-spatial positions. Consequently, these individuals may face unique social and epidemiological pressures, and the structure of the collective itself may alter contagion dynamics (e.g. super-spreader dynamics \cite{Lloyd-Smith2005}). This finding also has implications outside ecology. For example, in the design of collective systems, it implies that superior (up-to the threshold of $D=3$) individuals should focus on increased interaction with other individuals rather than increase their own exploration. As $D$ increases far beyond this threshold, the optimal strategy is for the distinct forager to engage in additional exploration (necessarily, as it cannot share more points with others), as shown by the decrease in $C_1$ after $D=3$ (Figure \ref{fig:unique_no_F}b). Beyond the threshold, the group-level information transfer plateaus (Figure \ref{fig:unique_no_F}d), implying no additional group-level gains from improving the ability of a superior agent. Therefore, in the design of collective systems (e.g. in the set up of robotic swarms), there is a limiting advantage to introducing new information gatherers which are particularly stronger than the rest of the group. Furthermore, this implies that distinct foragers in animal systems may take the most central socio-spatial position at the threshold of $D=3$, when the distinct forager engages in maximal sharing, and the corresponding variation in the social or epidemiological dynamics may be greatest for such foragers. While there is variation in the proportion of space shared by the distinct forager across the tested group sizes (with the forager sharing more of their uniquely known points in larger groups; $C_1$ is increasing with $n$), the qualitative shift in the dynamics of the optimal structure with $D$ after $D=3$ is consistently observed (Figure \ref{fig:unique_no_F}b), suggesting that this threshold may be of further theoretical interest.

\subsection{Resource-constrained environment}\label{sec:unique_forage}

\begin{figure}[t!]
    \centering
    \includegraphics[width=0.9\linewidth]{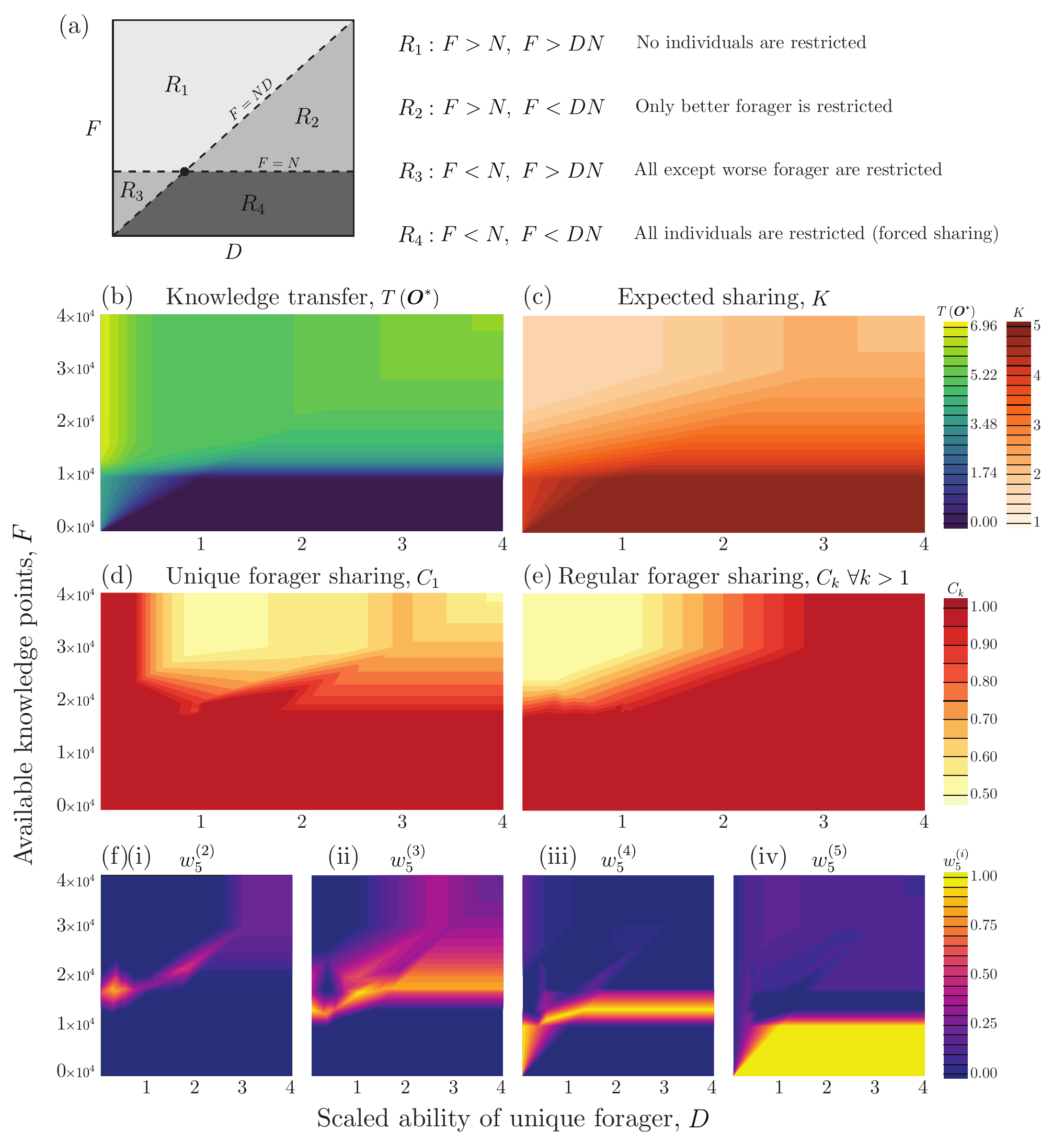}
    \caption{Results for the distinct forager scenario where $\boldsymbol{N}=(DN, N, \dots, N)$ with $D\in (0, 4]$, with foraging constraint varying between $F\in [0, 4\times 10^4]$, and for group size $n=5$. Specific values of $D$ and $F$ are obtained from 150 uniformly spaced points over their respective intervals, each rounded upwards. Shown are: (a) a schematic for the division of the $(D,F)$ parameter space into four regions, as outlined in the main text, (b) the value of the objective function at the optimal solution, $T(\boldsymbol{O}^*)$, (c) the expected number of individuals with knowledge of each point under the optimal structure, $K$, (d) the proportion of distinct foragers ($DN$)-many points also known by other individuals, $C_1$, (e) the proportion of regular individual's $N$-many points are also known by other individuals, $C_k$ (for $k>1$, fixed by problem symmetry), and (f) the relative orders of spatial overlap $w_5^{(i)}$ for (i) $i=2$, (ii) $i=3$, (iii) $i=4$ and (iv) $i=5$.}
    \label{fig:FD_contours}
\end{figure}

We find that the optimal spatial structure is dependent upon the combination of the relative ability of the distinct forager, $D$, and the resource availability, $F$. In particular, variation in measures of the optimal structure across the $(D,F)$ parameter space clearly separates the space into 4 regions, as defined in Figure \ref{fig:FD_contours}a. These regions, denoted by $R_1$, $R_2$, $R_3$ and $R_4$, are defined by which individuals are `restricted' (meaning whose effective knowledge is $F$, rather than their knowledge capacity $N_i$) by their environment for the given values of $(D, F)$. Particularly, the distinct forager is restricted by their environment when $F<DN$, and the regular individuals are restricted when $F<N$. In Region $R_1$ we typically observe stronger dependencies upon the value of $D$ (shown by the constant vertical regions which appear in many of the plots in Figure \ref{fig:FD_contours}), whereas in region $R_2$ we \textit{consistently} observe dependence only upon $F$ (as shown by the constant horizontal regions in Figure \ref{fig:FD_contours}). This is because the distinct forager is restricted to occupying $F$ many points, and there is no impact of variation of their actual ability $D$ within this region of parameter space. In region $R_3$, all recorded features vary monotonically with $D$, with the rate of variation being higher in more constrained systems (i.e. when $F$ is lower). Finally, within Region $R_4$, all individuals of the group have foraging abilities greater than that of their environment, so are forced into sharing all of the $F$-many points available (due to the assumption that individuals utilise their foraging abilities as far as their environment will allow). In this case, where complete sharing is forced, there is no knowledge transfer due to the lack of any unique knowledge (each point is known by all individuals and the spatial structure is trivial, i.e. $U_c(\boldsymbol{O})=0$ $\forall c$, $K=n$ and $w_n^{(n)}=1$).

As in the homogeneous scenario (Section \ref{sec: homo}), we observe patterns of interaction order selection in how the optimal spatial structure varies with $F$ (where increasing $F$ led to the optimal social structure consisting of sequentially lower-order interactions). In particular, surrounding the border of region \textbf{D} (where $w_n^{(n)}=1$) there is a region where $w_n^{(n-1)}$ peaks and $w_n^{(n)}$ falls, and surrounding this region there is another region where $w_n^{(n-2)}$ peaks and $w_n^{(n-1)}$ falls, and so on. This is exemplified in Figure \ref{fig:FD_contours}f for the case of $n=5$. However, not all responses of homogeneous systems to variation in resource abundance generalise to the distinct forager system. For example, there is an additional (local, not global) maximia in the value of $w_5^{(3)}$ for the homogeneous system (see Figures \ref{fig:homo_forage_ws}b and \ref{fig:FD_contours}f.ii), which is localised around $D=1$ in Figure \ref{fig:FD_contours}f.ii, and is therefore unique to groups with similarly skilled foragers. This suggests that there may be qualitative differences in how groups with homogeneous and non-homogeneous knowledge (acting under optimal collective processing schemes) may respond to the availability of resources in the environment. In particular, heterogeneous groups may distribute knowledge between subgroups of a different size to that of homogeneous groups, represented here through the different orders of space sharing observed. 

The shift in behaviour which was observed at $D=3$ in the resource-abundant scenario does not persist in the foraging constrained scenario, with the local maxima in both $K$ and $C_1$ (which characterised this shift in behaviour) disappearing at intermediate foraging constraints (compare Figures \ref{fig:FD_contours}c and \ref{fig:FD_contours}d). Therefore in animal systems the socio-spatial impact of distinct foragers upon the group may be less significant in resource-constrained scenarios, as the effective benefit of increased foraging ability is reduced.

\section{Heterogeneous knowledge scenario}\label{sec: hetero}

In this scenario we consider the impact of general heterogeneity in individual capacities for knowledge upon the optimal spatial structure of the group. The amount of heterogeneity is described through variation in the parameter $\sigma$, as outlined in Section \ref{sec: scenarios}. We again differentiate between resource-abundant and resource-constrained scenarios.  

\subsection{Resource-abundant environment}

\begin{figure}[t!]
    \centering
    \includegraphics[width=\linewidth]{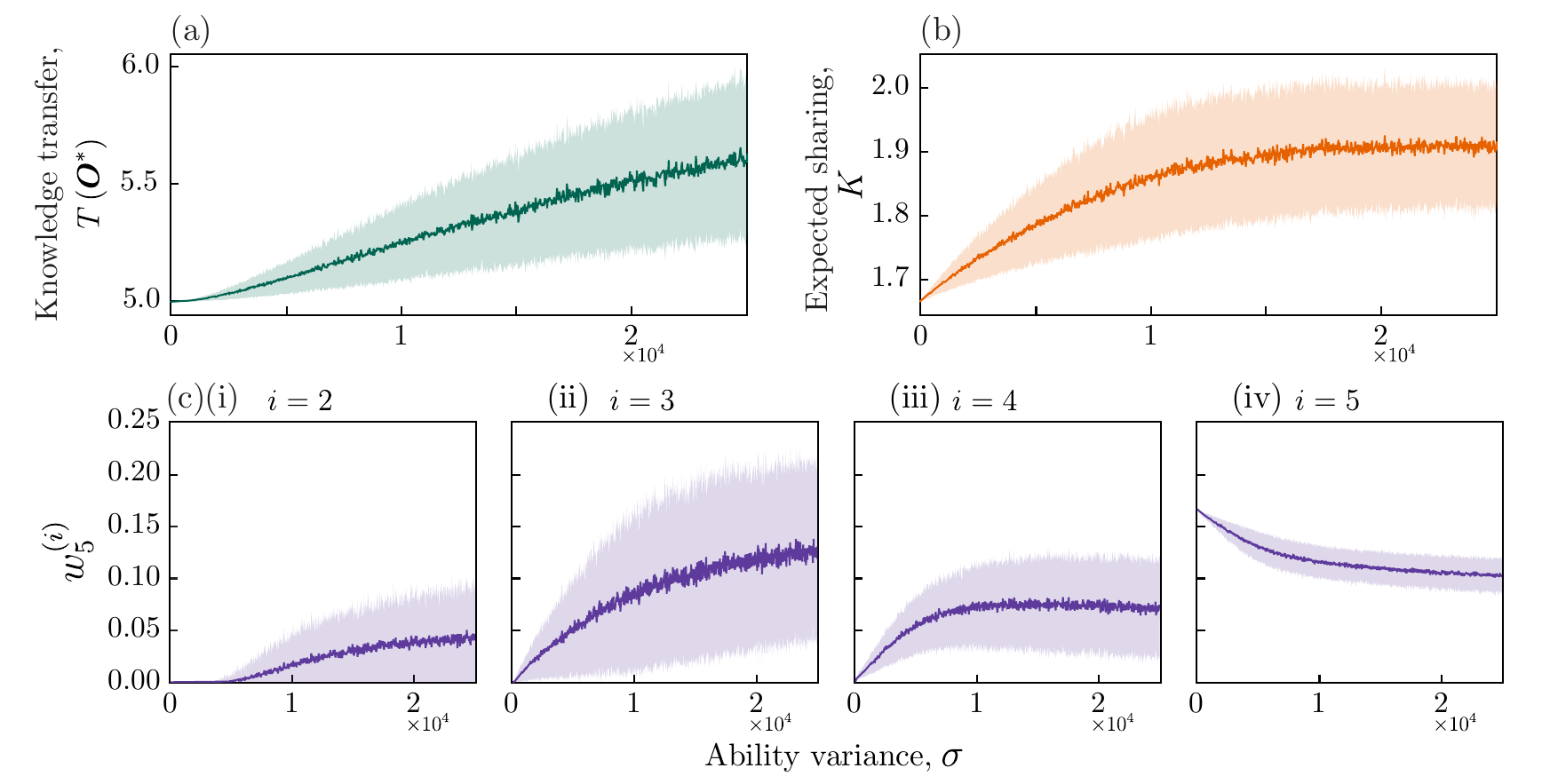}
    \caption{Results from our general heterogeneity scenario in the resource-abundant case with $\sigma\in [0, 2.5\times 10^4]$, mean foraging ability $N=10^4$ and group size $n=5$. Specific values of $\sigma$ are obtained from 1000 uniformly spaced points in this interval rounded upwards. Shown are: (a) the value of the objective function at the optimal solution, $T(\boldsymbol{O}^*)$, (b) the expected number of individuals with knowledge of each point under the optimal structure, $K$, (c) the relative orders of spatial overlap, $w_5^{(i)}$, for (i) $i=2$, (ii) $i=3$, (iii) $i=4$ and (iv) $i=5$. In each plot, solid lines are the average value across each of the 250 Monte Carlo simulations for each value of $\sigma$, and shaded regions show one standard deviation from this point.}
    \label{fig:general_hetro_noF}
\end{figure}

With increasing heterogeneity (increasing $\sigma$), the proportion of highest-order space sharing, $w_n^{(n)}$, falls and there is a rise in the proportion of each lower-order sharing, $w_n^{(i)}$ for all $i=2,\dots, n-1$ (Figure \ref{fig:general_hetro_noF}c). This suggests that the most centralised spatial structures appear in groups with homogeneous foraging abilities. Ecologically, this implies that homogeneous groups may exhibit greater group cohesion whenever they are organised for optimal collective knowledge processing as all individuals have intersection in their core areas. This may facilitate the emergence of other group social traits related to high group attendance, such as common sleeping areas. Also as $\sigma$ increases, there is a steady increase in both the group-level knowledge transfer, $T$ (Figure \ref{fig:general_hetro_noF}a), and the expected point knowledge $K$ (Figure \ref{fig:general_hetro_noF}b). In groups with overdispersed knowledge levels, where $\sigma > N$, each characteristic of the optimal spatial structure (captured within this study) levels off to a fixed value (see Figure S2 for figures with a wider range of $\sigma$ values). This suggests that the differences in spatial structures between two highly heterogeneous groups acting under optimal regimes may be limited, even if one group has a relatively higher degree of variation in their abilities. Particularly, in nature we my expect groups with high variation and groups with extremely high variation in foraging abilities to exhibit similar spatial structures when placed in similarly resource abundant environments (when group-level optimality is important). These patterns are consistent across tested group sizes. 

\subsection{Resource-constrained environment}
\begin{figure}[t!]
    \centering
    \includegraphics[width=\linewidth]{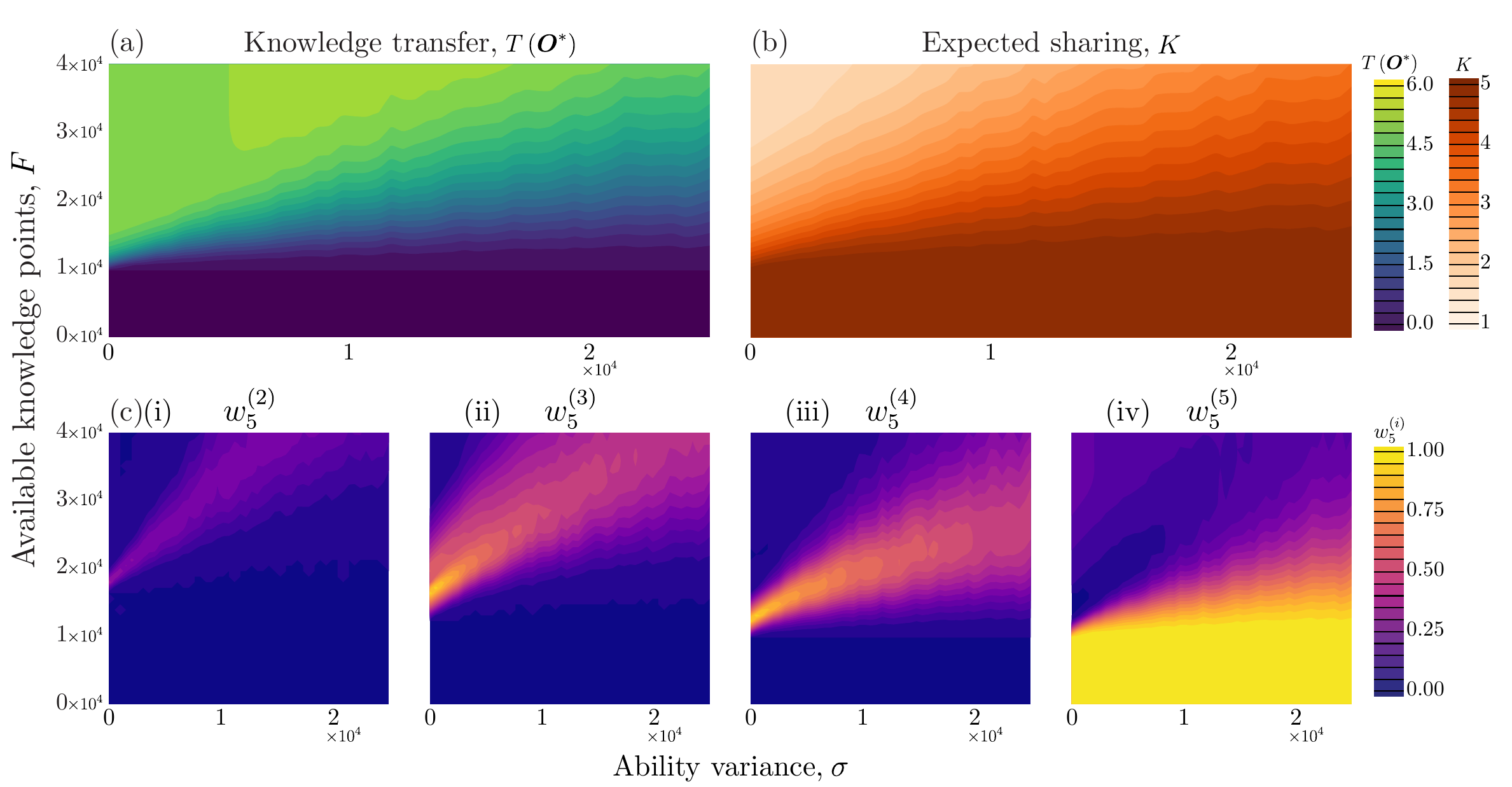}
    \caption{Results from our general heterogeneity scenario with $\sigma\in [0, 2.5\times 10^4]$, mean foraging ability $N=10^4$, foraging constraint varying between $F\in [0, 4\times 10^4]$ and for group size $n=5$. Specific values of $\sigma$ and $F$ are obtained from 50 uniformly spaced points over their respective intervals, each rounded upwards. Shown are: (a) the value of the objective function at the optimal solution, $T(\boldsymbol{O})$, (b) the expected number of individuals with knowledge of each point under the optimal structure, $K$, (c) the relative orders of spatial overlap $w_5^{(i)}$ for (i) $i=2$, (ii) $i=3$, (iii) $i=4$ and (iv) $i=5$. In each contour plot, every value is the average value across each of the 250 Monte Carlo simulations for the corresponding value of $\sigma$ and $F$.}
    \label{fig:hetro_contours}
\end{figure}

We find that many values of $F$ reverse the relationship found in the resource-abundant case: the proportion of highest-order sharing, $w_n^{(n)}$, can increase with $\sigma$, while the proportion of lower-order sharing, $w_n^{(i)}$, can decrease with increasing within-group variation in abilities $\sigma$ for $i=2,\dots, n-1$. Therefore, groups with greater variation in individual information capacities may have a centralised spatial structure if they are placed in an environment with constrained resources, in contrast to the scenario where resources are abundant. Some values of $F$ can also lead to $w_n^{(i)}$ growing initially with $\sigma$ and then decreasing after some peak value (e.g. $w_5^{(3)}$ at $F=20000$ in Figure \ref{fig:hetro_contours}c.ii). This change in behaviour upon the introduction of foraging constraints may be because with higher $\sigma$ there is a higher likelihood that some individual abilities, $N_i$, are significantly higher than average (due to the wider distribution from which $\boldsymbol{N}$ is drawn from), and these individuals may be more harshly constrained by lack of resources in their environment. This is evidenced by the fact that as $\sigma$ grows, the group-level knowledge transfer begins to fall towards 0 at higher values of $F$ (Figure \ref{fig:hetro_contours}a). Hence, in constrained environments the heterogeneity of the group may actually be restricted as within-group variation in foraging abilities, $\sigma$, increases (as more individuals may be limited by poor resource availability and therefore have equal effective knowledge of $N_i'=F$). This is evidenced through our consideration of a normalised case, where the mean of $\boldsymbol{N}$ is fixed across each $\sigma$ value (limiting how significant the outliers in foraging abilities can be) in which we did not observe as significant variation when varying $F$ (Figure S3). Note that this normalisation did not have an impact upon our results for the resource-abundant scenario (Figure S4). 

We continue to observe patterns of interaction order selection with $F$ in this scenario, as in Sections \ref{sec:Homo_withF} and \ref{sec:unique_forage}. Particularly, for each $\sigma$ value, there are peaks in the values of $w_n^{(i)}$, where $i$ is increasing sequentially with $F$ (and this pattern persists when normalising $\boldsymbol{N}$). This is represented in Figure \ref{fig:hetro_contours}c, by the streaks of maximal values of $w_n^{(i)}$ which are decreasing in `gradient' with $i$. The fact that this pattern was observed consistently across our different scenarios implies that resource availability can select for a specific order of interaction in a diversity of systems. These `streaks' of higher values in the heterogeneous population only reach global maximality (i.e. $w_n^{(i)}=1$) for small values of $\sigma$, when abilities are more homogeneous. Therefore, more uniform/ordered spatial structures (where the order of interactions is fixed throughout) may be a unique property in the optimal spatial structure of groups with more homogeneous information gathering capacities. This highlights how within-group variation in foraging abilities may be a key driver of spatial complexity in a variety of natural systems, and suggests that the movement algorithms of artificial systems should be designed with the amount of device heterogeneity (e.g. in movement capabilities, storage capacities) explicitly accounted for.

\section{Conclusions}\label{sec: conclusions}

Our model yielded important insights into how collectives can organise their spatial structures to promote optimal sharing of knowledge, which is of strong relevance across domains in heterogeneous environments. Our three scenarios explored how the optimal spatial structure of a collective foraging group may vary according to differences in individual abilities, and how these structures may respond differently to variation in the resource abundance of their environment.

One key finding was that the distribution of knowledge in optimally structured groups was heavily influenced by the level of resource availability (represented in this study through the parameter $F$). Particularly, we consistently observed that the value of $F$ promoted specific orders of space sharing in the optimal structure of the group, which may be interpreted as certain subgroup sizes. In both lightly and heavily constrained environments our results suggested that we would expect commonly-shared foraging points within the core areas of all group members, in line with the information centre hypothesis \cite{Evans2016} for ecological systems. However, systems which are heterogeneous in their information gathering abilities and placed in environments with intermediate resource levels were found to instead have their optimal spatial structures dominated by lower-order spatial overlaps. This implies that we may expect a relatively worse support for the information centre hypothesis in such systems and a better support for other, less centralized foraging strategies. If we interpret prominence of lower-order interactions as indicative of a fission-fusion type society (primarily observed in heterogeneous environments \cite{Aureli2008, Dunbar1988} and characterised by reduced group cohesion and by a high fluidity in subgroup composition \cite{Couzin2009}), then our model suggests that intermediate resource availability or heterogeneity in foraging abilities can give rise to stronger fission-fusion dynamics. This result aligns with previous theory relating fission-fusion dynamics and foraging behaviour \cite{Ramos-Fernandez2006}. It is interesting that we reach this similar conclusion using a different framework (based on knowledge sharing rather than individual foraging).

Using specific orders of spatial interaction allowed collectives to maintain high levels of knowledge sharing in constrained environments, but this organisation caused each individual to maintain fewer uniquely known foraging points (quantified here by $C_k$), potentially representing an individual-level cost of the spatial organisation of the collective \cite{Lee2016}. These potential costs were higher in larger groups (for fixed levels of resource abundance, $F$), with larger groups also exhibiting more crowding around foraging sites (which, in animal systems, may devalue the knowledge of these points by increasing competition). This has implications in the evolutionary ecology of group size. In particular, in sufficiently resource abundant environments, we observed quadratic-like growth in the objective function value with group size, $n$, such that an increase in group size translates to a linear benefit to each individual (\textit{on average}, since benefits could be spread unequally across the group). Resource availability of the environment might then select for a particular group size (in agreement with pre-existing theory; \cite{Seiler2020, Gibbs1987}), by determining the individual-level costs to a particular structure compared to the increased benefit from the group-level (both of which should increase with group size). This is in agreement with pre-existing theory that intermediately sized groups have energetically optimal space-use strategies \cite{Markham2015}, suggesting that the pressure to organise to maximise information transfer can promote intermediate group sizes. In artificial systems, these individual-level costs could represent the financial costs of a single agent/robot \cite{Schroeder2019}, and expected resource availability could therefore help to inform budgeting strategies in collective systems design. Further analytical work in this framework could help bypass computational issues regarding dimensionality, and yield a more precise synthesis of the role of group size in collective intelligence systems.

Our model framework is general and is designed to be applicable to a broad range of collective systems. There are a variety of natural extensions to this model which may increase applicability. An interesting direction would be the inclusion of non-independent movement of individuals. Ecologically, the assumption of independence implies that we are modelling a system where either group cohesiveness has not yet formed, where information gathering is an inherently solitary task, or where there is high de-synchronization of needs and movement motivations (e.g. systems with high fission-fusion dynamics and low predation pressure; \cite{Aureli2008}). By explicitly tracking each potential foraging point, one could implement following behaviours, which are well-established empirically \cite{Palacios-Romo2019}. A similar system to this optimisation framework but with an agent-based framing could allow the model to represent and compare different movement algorithms for robot swarms \cite{Tarapore2020} in variable environments and consider variation in the number of individuals - possibly helping to inform systems design in novel applications of swarm technology. There is a rich variety of other possible future directions, including: heterogeneity in resource values, explicit temporal variation in resource availability and multi-level optimisation, which may all help to further explain the dramatic variation in spatial structures observed in nature and help to inform collective systems design. 

Collectively, our paper highlights how resource availability and variation in foraging capacities can influence the optimal spatial structures of collective foraging or information gathering systems. Given the rising prevalence of robot swarms in human applications, and rapid global change driving variation in food abundance for animal systems, such understanding could shape much needed theoretical and empirical research across diverse contexts. Our work, therefore, exemplifies the importance of considering higher-order spatial interactions in studies of natural and artificial collectives. Further examination of higher-order spatial structure could generate insights of broad ecological relevance (e.g. animal colonies), with potential applications for the bio-inspired design of artificial collective systems (e.g. swarm robotics systems). 

\clearpage
\printbibliography
\section*{Code availability}
The underlying code for this study is available at: \url{https://github.com/rswalker-eco/Spatial_topological_optimisation.git}

\section*{Acknowledgements}
RSW was supported by the EPSRC Centre for Doctoral Training in Mathematical Modelling, Analysis and Computation (MAC-MIGS) funded by the UK Engineering and Physical Sciences Research Council (grant EP/S023291/1). GRF partly conducted this work during a sabbatical stay at the Global Research Centre for Diverse Intelligences at the University of St. Andrews, being supported by a PASPA-DGAPA grant from the Universidad Nacional Autónoma de México. SESA was supported by a postdoctoral grant (\textit{Estancias posdoctorales por México}) from the Secretaría de Ciencia, Humanidades, Tecnología e Innovación (SECIHTI). XO was supported by the BBSRC EEID research grant BB/V00378X/1. MJS was supported by the Royal Society University Research Fellowship URF/R1/221800. We would thank Nina H. Fefferman for her comments on the conceptualisation of the model. 

\section*{CRediT author contributions} 
\textbf{RSW:} Conceptualisation, Methodology, Software, Validation, Formal analysis, Writing - Original Draft, Writing - Review \& Editing, Visualisation \textbf{GRF:} Conceptualisation, Writing - Review \& Editing, Supervision \textbf{DB:} Conceptualisation, Writing - Review \& Editing \textbf{SESA:} Writing - Review \& Editing, Visualisation \textbf{XO:} Methodology, Writing - Review \& Editing \textbf{MJS:}  Conceptualisation, Writing - Review \& Editing, Supervision

\end{document}


\maketitle

\section{Additional methods details}
Throughout, we maintain the same notation as the main text. Individuals in a group of size $n$ are arbitrarily enumerated from 1 to $n$. We define $\mathcal{N}=2^n-n-1$ as the number of possible subgroups (meaning subsets of the group consisting of strictly more than 1 individual), $\mathcal{X}$ as the collection of these sets and $I\colon \mathcal{X} \to [\mathcal{N}]$ as the enumeration map of $\mathcal{X}$ (as defined in the main text). We also have the definitions of the sets $S(c)$ and $B(c)$ for $c\in \left[\mathcal{N}\right]$ as defined in Equations \eqref{eq: def S} and \eqref{eq: def B} respectively.
\begin{align}
&S(c) = \left\{i\in \left[\mathcal{N} \right] : I^{-1}(c)\subseteq I^{-1}(i)\right\}\label{eq: def S}  \\
&B(c) = \left\{i\in \left[\mathcal{N} \right] : I^{-1}(i) \subset I^{-1}(c)\right\}\label{eq: def B}
\end{align}
We defined the problem variables as $\boldsymbol{O}= \left(O_i\right)_{i\in [\mathcal{N}]}$ where $O_i$ gives the number of points known uniquely by the $i$-th subgroup in $\mathcal{X}$ under the enumeration $I$.

\subsection{Explicit form of objective function coefficients}\label{sec: coefficients}
Let $\boldsymbol{N}\in \mathbb{Z}^n$ be the vector of individual knowledge of foraging points. In the main document, we defined the objective function as 
\begin{align}
    T(\boldsymbol{O}) &= \sum_{c=1}^\mathcal{N} P_c(\boldsymbol{O})U_c({\boldsymbol{O}}) \nonumber \\
    & = \sum_{c=1}^\mathcal{N} \frac{\sum_{a\in S(c)} O_a}{\prod_ {k\in I^{-1}(c)}N_k'}\left(
\sum_{k\in I^{-1}(c)}N_k' - f(c)\sum_{a\in S(c)}O_a-\sum_{a\in B(c)}(f(c)-1)O_a
    \right)\label{eq: T full form} \\
    & = \boldsymbol{l}^t\boldsymbol{O}+\frac{1}{2}\boldsymbol{O}^tM\boldsymbol{O} \label{eq: T as quad}
\end{align}
for some $\boldsymbol{l}=(l_i)\in \mathbb{R}^{\mathcal{N}}$ and $M=(M_{ij})\in\mathbb{R}^{\mathcal{N}\times\mathcal{N}}$ is symmetric. In this section we derive the specific entries of $\boldsymbol{l}$ and $M$ by grouping together terms in Equation \eqref{eq: T full form} into the form of Equation \eqref{eq: T as quad}. We express $T(\boldsymbol{O})$ in terms of its linear part, $L(\boldsymbol{O})$, and quadratic part $Q(\boldsymbol{O})$, such that 
\begin{equation*}
    T(\boldsymbol{O}) = L(\boldsymbol{O}) + Q(\boldsymbol{O})
\end{equation*}
where $L$ and $Q$ are defined explicitly in Equations \eqref{eq: lin part} and \eqref{eq: quad part} respectively.
\begin{align}
    &L\left(\boldsymbol{O}\right) = \sum_{c=1}^{\mathcal{N}}\left[\frac{\sum_{k\in I^{-1}(c)}N_k'}{\prod_{k\in I^{-1}(c)}N_k'}\sum_{a\in S(c)}O_a\right] \label{eq: lin part}\\
    &Q\left(\boldsymbol{O}\right)=-\sum_{c=1}^{\mathcal{N}}\left[\frac{1}{\prod_{k\in I^{-1}(c)}N_k'} \left(f(c)\sum_{a_1, a_2\in S(c)}O_{a_1}O_{a_2}+ \sum_{\substack{a_1 \in B(c)\\ a_2 \in S(c)}} (f\left(a_1\right)-1)O_{a_1}O_{a_2}\right)\right] \label{eq: quad part}
\end{align}
We can then determine $l$ by simplifying $L(\boldsymbol{O})$ and $M$ by simplifying $Q(\boldsymbol{O})$. 

We begin by collecting the linear coefficients. There is a contribution from the first sum in $L$ (over $c$) to the coefficient of $O_i$ (for each $i\in [\mathcal{N}]$) only for the values of $c\in \left[\mathcal{N}\right]$ with $i\in S(c)$. This motivates the definition of the set $J(i)$ as 
\begin{equation*}
    J(i) =\left\{c\in \left[\mathcal{N}\right] : i\in S(c)\right\}.
\end{equation*}
We then write the coefficient of $O_i$ in $L(\boldsymbol{O})$, which is exactly the $i$-th entry of $\boldsymbol{l}$, as 
\begin{equation*}
    l_i = \sum_{c\in J(i)} \left[\frac{\sum_{k\in I^{-1}(c)}N_k'}{\prod_{k\in I^{-1}(c)}N_k'}\right]
\end{equation*}

We now collect the quadratic coefficients, to determine $M$. We consider the two components of $Q(\boldsymbol{O})$ separately:
\begin{equation*}
    Q(\boldsymbol{O}) = Q_1(\boldsymbol{O}) + Q_2(\boldsymbol{O})
\end{equation*}
where $Q_1(\boldsymbol{O})$ and $Q_2(\boldsymbol{O})$ are defined in Equations \eqref{eq: Q1} and \eqref{eq: Q2} respectively. 
\begin{align}
    &Q_1\left(\boldsymbol{O}\right) = -\sum_{c=1}^{\mathcal{N}}\left[\frac{1}{\prod_{k\in I^{-1}(c)}N_k'} \left(f(c)\sum_{a_1, a_2\in S(c)}O_{a_1}O_{a_2}\right)\right]\label{eq: Q1}\\
    & Q_2\left(\boldsymbol{O}\right) = -\sum_{c=1}^{\mathcal{N}}\left[\frac{1}{\prod_{k\in I^{-1}(c)}N_k'} \left(\sum_{\substack{a_1 \in B(c)\\ a_2 \in S(c)}} (f\left(a_1\right)-1)O_{a_1}O_{a_2}\right) \right] \label{eq: Q2}
\end{align}
Let $i,j\in [\mathcal{N}]$. For the sum in $Q_1(\boldsymbol{O})$, there is a contribution in to the coefficient of $O_iO_j$ only for the values of $c$ with $i\in S(c)$ and $j\in S(c)$, meaning the $c$ values such that $I^{-1}(c)\subseteq I^{-1}(i)$ and $I^{-1}(c)\subseteq I^{-1}(j)$. Collectively this implies that $I^{-1}(c)\subseteq I^{-1}(i)\cap I^{-1}(j)$ must hold. This motivates the definition of another set $V(i,j)$, which is given by 
\begin{equation*}
    V(i,j) = \left\{c: I^{-1}(c)\subseteq I^{-1}(i)\cap I^{-1}(j) \right\}.
\end{equation*}
For each $c$ in this set, we will get a contribution of $2f(c)$ to the coefficient of $O_iO_j$ in $Q_1$ if $i\neq j$ (the scaling of $2$ comes from the fact that the \textit{unordered} pair $(i,j)$ appears twice in the sum, from the cases $a_1=i$, $a_2=j$ and $a_1=j$, $a_2=i$) and a contribution of $f(c)$ if $i=j$. This allows us to write the coefficient of $O_iO_j$ in $T$ from $Q_1$, denoted $\operatorname{Coeff}_1(O_iO_j)$, as
\begin{equation*}
    \operatorname{Coeff}_1(O_iO_j) = \begin{cases}
         -\sum_{c\in V(i,j)}\frac{2f(c)}{\prod_{k\in I^{-1}(c)}N_k'}\text{, if } i\neq j \\ -\sum_{c\in V(i,j)}\frac{f(c)}{\prod_{k\in I^{-1}(c)}N_k'}\text{, if } i=j
    \end{cases}
\end{equation*}
Now, for from $Q_2(\boldsymbol{O})$ there will be a contribution to the coefficient of $O_iO_j$ only if $i\in S(c)$ and $j\in B(c)$ (or the other way around, which is handled by symmetry of $M$). This is equivalent to the contributing $c$ values being those such that $I^{-1}(c)\subseteq I^{-1}(i)$ and $I^{-1}(j)\subset I^{-1}(c)$. Note that this condition implies, by transitivity, that there will only be a contribution to $O_iO_j$ if $I^{-1}(j) \subset I^{-1}(i)$, which implies that $i>j$. To collect all $c$ values satisfying these two conditions, we define a further set $Z(i:j)$ such that 
\begin{equation*}
    Z(i: j) = \left\{c : I^{-1}(j)\subset I^{-1}(c)\subseteq I^{-1}(i) \right\}
\end{equation*}
where the colon notation is used to emphasise that $Z(i:j)\neq Z(j:i)$. This construction allows us to write the contribution from $Q_2$ to the coefficient of $O_iO_j$ for $i>j$ as 
\begin{equation*}
    \operatorname{Coeff_2}(O_iO_j) = \begin{cases}
        -\sum_{c\in Z(i: j)}\frac{g(j)}{\prod_{k\in I^{-1}(c)}N_k'}\text{, if } I^{-1}(j)\subset I^{-1}(i) \\
        0, \text{otherwise}
        \end{cases}
\end{equation*}
The case for $i<j$ is handled by symmetry. Finally, the coefficient of $O_iO_j$ will be given by $\operatorname{Coeff_1}(O_iO_j)+\operatorname{Coeff_2}(O_iO_j)$. This collection of coefficients defines the entries of the matrix $M$ as
\begin{equation}\label{eq: m coefficients}
    M_{ij} = \begin{cases}
        2\operatorname{Coeff}(O_iO_j)\text{, if } i=j \\
    \operatorname{Coeff}(O_iO_j)\text{, otherwise }
    \end{cases}
\end{equation}
where the sign is reversed because $Q(\boldsymbol{O})$ has a coefficient of $-1$ in $T(\boldsymbol{O})$. Note that all entries of $M_{ij}$ will be non-positive. Note that since we double the diagonal terms (where $i=j$) in Equation \eqref{eq: m coefficients}, the effective contribution from $\operatorname{coeff}_1(O_iO_j)$ to $M$ is always 
\begin{equation*}
    m_{ij, 1} \vcentcolon= -\sum_{c\in V(i,j)}\frac{2f(c)}{\prod_{k\in I^{-1}(c)}N_k'}
\end{equation*} 
The residual part of $M_{ij}$ which is unaccounted for by $m_{ij, 1}$ comes from $\operatorname{coeff}_2(O_iO_j)$, and we denote this by $m_{ij, 2}$ such that $M_{ij} = m_{ij, 1} + m_{ij, 2}$.

\subsection{Explicit form of inequality constraint coefficients}
As described in the main text, the inequality constraints of the problem are given in Equations \eqref{const: non-neg}, \eqref{const: no-overshare}, and \eqref{const: resources}, where $\tilde{S}(k)$ is defined in Equation \eqref{eq: S_tilde}.
\begin{align}
    &\boldsymbol{O}\geq \boldsymbol{0} \label{const: non-neg}\\
    & \sum_{i\in \tilde{S}(k)}O_i\leq N_k', \text{ for all } k\in [n] \label{const: no-overshare}\\
    &\sum_{i=1}^nN_i' - \sum_{c=1}^\mathcal{N}(f(c)-1)O_c \leq F\label{const: resources} \\
    &\tilde{S}(k) = \left\{i\in[\mathcal{N}]: \{k\}\subset I^{-1}(i)\right\}\label{eq: S_tilde}
\end{align}
The constraints in Equations \ref{const: no-overshare} and \ref{const: resources} are both linear in $\boldsymbol{O}$, and therefore can be collectively described by: 
\begin{equation*}
    G\boldsymbol{O}\leq h
\end{equation*}
for some $G=(G_{ij})\in \mathbb{R}^{(n+1)\times \mathcal{N}}$ and $h=(h_i)\in\mathbb{R}^{n+1}$. 

The first $n$ rows of $G$ and $h$ are defined by the constraint in Equation \eqref{const: no-overshare}. The entries in these rows of $G$, denoted $G_{ij}$ for $i\in [n]$ and $j\in [\mathcal{N}]$, will only be 0 or 1, taking value 1 only if $j\in\tilde{S}(i)$. The corresponding entry in the $i$-th row of $h$ will be $N_i'$. The final row of $G$ is determined by the foraging resource constraint, Equation \eqref{const: resources}. The $j$-th entry in this row will be precisely the coefficient of $O_j$ in Equation \eqref{const: resources}, which is $1-f(j)$. The final entry of $h$ will be:
\begin{equation*}
    h_{n+1} = F - \sum_{i=1}^nN_i'.
\end{equation*}
In summary, for $i\in [n+1]$ and $j\in [\mathcal{N}]$: 
\begin{align*}
    & G_{ij} = \begin{cases}
        1 \text{, if } \hspace{4pt} i\in [n], j\in\tilde{S}(i) \\
        0 \text{, if } \hspace{4pt} i\in [n], j\notin\tilde{S}(i) \\
        1-f(j) \text{, if } \hspace{4pt} i = n+1  
    \end{cases},\\
    & h_i = \begin{cases}
        N_i' \text{, if} \hspace{4pt} i\in [n] \\
        F - \sum_{i=1}^nN_i' \text{, if} \hspace{4pt} i=n+1
    \end{cases}.
\end{align*}

\subsection{Problem feasibility}
We show here that the optimisation problem given in Equations (2.3)-(2.4) of the main text is feasible whenever $F\geq 0$. For this, we show that the region of feasible solutions,
\begin{equation*}
    \Omega = \{\boldsymbol{O}\in \mathbb{Z}^\mathcal{N}: \boldsymbol{O}\geq\boldsymbol{0}, G\boldsymbol{O \leq h}\}
\end{equation*}
is non-empty and finite for each $F\geq 0$ and each $\boldsymbol{N}\geq 0$. To show that $\Omega$ is non-empty we construct the value of $\boldsymbol{O}$ which represents the case of maximal sharing - where all individuals are sharing as much of their known points as possible with others. We then show that this solution satisfies the problem constraints, in the forms provided in Equations \eqref{const: non-neg}, \eqref{const: no-overshare} and \eqref{const: resources}.

Write $\boldsymbol{N} = (N_i)_{i\in[n]}$. Then as in the main text, for a given resource level, $F$, the effective knowledge of the $i$-th individual is 
\begin{equation}\label{eq: effective knowledge}
    N_i'=\min\{N_i, F\}. 
\end{equation}
Without loss of generality, suppose that individuals are indexed such that $N_i\leq N_j$ for $i<j$. Also suppose that the collection $\{N_1, N_2,\dots, N_n\}$ has $m$-many unique values, and denote these by $H_1< \dots < H_m$ for $m\leq n$. Let the number of individuals which have effective knowledge values of $H_k$ be denoted $\varphi(k)$ for $k\in[m]$. We then define $\Phi(k)$ as the number of individuals having effective knowledge no more than $H_k$, given by: 
\begin{equation*}
    \Phi(k) = \sum_{i=1}^k \varphi(i)
\end{equation*}
for $k\in[m]$. We also define $\Psi(k)$ as the collection of individuals with indices strictly greater than $\Psi(k)$ for $k\in [m-1]$. Particularly:
\begin{equation*}
    \Psi(k) = \{\Phi(k)+1, \dots, n\}
\end{equation*}

\begin{figure}[t!]
    \centering
    \includegraphics[width=0.65\linewidth]{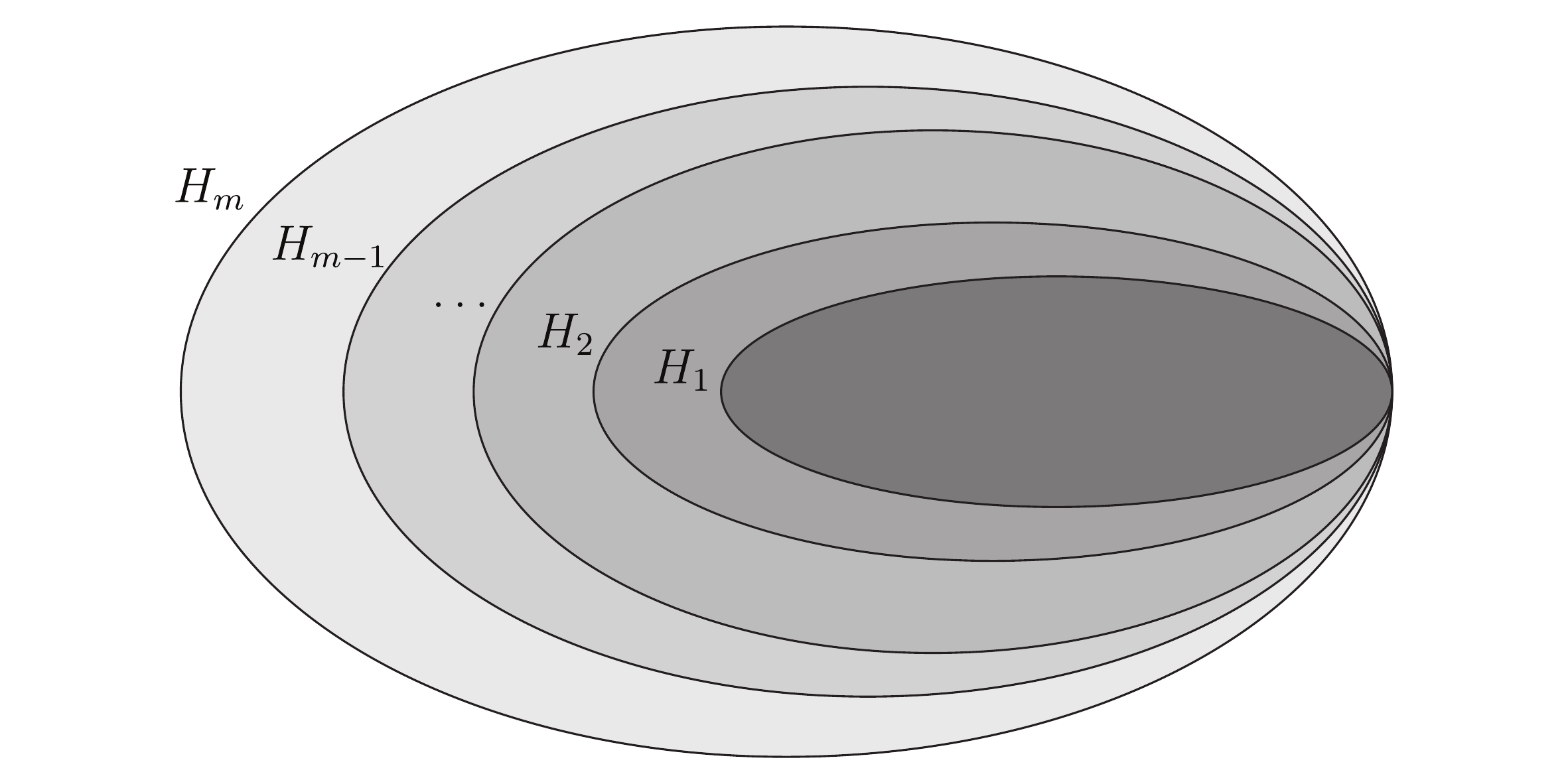}
    \caption{Nested structure of maximal sharing. Such structure is always feasible, and represents the scenario with minimal unique knowledge for fixed values of $F$ and $\boldsymbol{N}$.}
    \label{fig:maximal_sharing}
\end{figure}

These constructions allow us to represent the value of $\boldsymbol{O}\in\mathbb{Z}$ corresponding to the maximal sharing of foraging points. The particular structure we describe is nested, as visualised in Figure \ref{fig:maximal_sharing}. 

All individuals can share $H_1$-many points, since all effective abilities are less than or equal to $H_1$. Therefore, $O_\mathcal{N}=H_1$ in the structure corresponding to maximal sharing. Then all individuals except those with effective abilities of $H_1$, represented by $\Psi(1)$, can share $H_2$-many points. Although, $H_1$-many of those points are also shared with the rest of the group, so the number of points known uniquely by that subgroup is $H_2-H_1>0$. Similarly, the subgroup $\Psi(2)$ can share $H_3$-many points, but $H_2$ of those are shared with others, so the corresponding uniquely known number of points is $H_3-H_2$. Generalising this argument, our proposed feasible point is given by $\boldsymbol{O}=\{O_c\}_{c\in[\mathcal{N}]}$, where:
\begin{equation*}
    O_c = \begin{cases}
        H_1\text{, if} \hspace{4pt} c = \mathcal{N}\\
        H_{k+1}-H_{k}\text{, if} \hspace{4pt} c=I\left(\Psi(k)\right) \text{  for some } k\in [m-1]\\
        0\text{, otherwise}
    \end{cases}
\end{equation*}

It remains to show that this point is feasible. Since $H_1, \dots, H_m$ is an increasing sequence, it is clear that $O_c\geq 0$ for all $c\in [\mathcal{N}]$, so the constraint in Equation \eqref{const: non-neg} is satisfied. Also, by construction, each of the constraints in Equation \eqref{const: no-overshare} are satisfied. We need to verify that the resource availability constraint (Equation \eqref{const: resources}) is satisfied. For this, first observe that 
\begin{equation*}
    \sum_{i=1}^nN_i' = \sum_{k=1}^m \varphi(k)H_k.
\end{equation*}
Then, noting that $f(I(\Psi(k)))=n-\Phi(k)$ (since $\Psi(k)$ only contains individuals with indices strictly greater than $\Phi(k)$), we observe that the sum in the resource availability constraint,
\begin{equation}\label{eq: forage_constraint_maximal}
    \sum_{c=1}^\mathcal{N}O_c(f(c)-1) = (n-1)H_1 + \sum_{k=1}^m (H_{k+1}-H_k)(n-\Phi(k)-1),
\end{equation}
has a telescoping structure. Particularly, for $k=2,\dots m-1$, the coefficient of $H_k$ is determined by the following two consecutive terms in the sum:
\begin{equation*}
    (H_{k+1}-H_k)(n-\Phi(k)-1) + (H_{k}-H_{k-1})(n-\Phi(k-1)-1).
\end{equation*}
Using the fact that that $\Phi(k) = \Phi(k-1)+\varphi(k)$, this expression simplifies to 
\begin{equation*}
    H_{k+1}(n-\Phi(k)-1) + H_{k}\varphi(k) - H_{k-1}(n-\Phi(k-1)-1).
\end{equation*}
For $k=1$, the coefficient is determined by both the $(n-1)H_1$ term and the $k=1$ term in Equation \eqref{eq: forage_constraint_maximal}, yielding
\begin{align*}
    (n-1)H_1+(H_2-H_1)(n-\Phi(1)-1) &= nH_1+(H_2-H_1)(n-\varphi(1)-1) \\
    & = \varphi(1)H_1 + (n-\varphi(1)-1)H_2
\end{align*}
using the fact that $\Phi(1)=\varphi(1)$ by definition. Then the coefficient of $H_m$ is given by the final term in the sum in Equation \eqref{eq: forage_constraint_maximal}, only. Particularly, from the term 
\begin{equation*}
    (H_m-H_{m-1})(n-\Phi(m-1)-1) = (\varphi(m)H_m - H_m) - H_{m-1}(n-\Phi(m-1)-1)
\end{equation*}
where here we use the fact that $n-\Phi(m-1)=\varphi(m-1)$. We conclude that: 
\begin{equation*}
    \sum_{c=1}^\mathcal{N}O_c(f(c)-1) = \left(\sum_{k=1}^m H_k\varphi(k)\right) - H_m.
\end{equation*}
Therefore, we have that: 
\begin{align*}
    \sum_{i=1}^\mathcal{N}N'_i-\sum_{c=1}^\mathcal{N}O_c(f(c)-1) &= \sum_{k=1}^m H_k\varphi(k) - \left(\left(\sum_{k=1}^m H_k\varphi(k)\right) - H_m \right) \\ 
    & = H_m\\
    & = N_n' \\
    &\leq F
\end{align*}
where the final inequality follows automatically from Equation \eqref{eq: effective knowledge}. Therefore, $\boldsymbol{O}\in \Omega$, and $\boldsymbol{O}\neq \emptyset$. 

It remains to show that $\Omega$ is finite. We observe that $\Omega\subseteq [0, N_n]^n$, since any $O_i>N_n$ would automatically break at least one of the assumptions in Equation \eqref{const: no-overshare}. Thus, $\Omega$ is bounded. Therefore, since $\Omega\subset \mathbb{Z}$ by definition, $\Omega$ must be finite as a bounded set of integers. This implies that the optimisation problem given in Equations (2.3)-(2.4) of the main text is feasible, since $\Omega$ is non-empty and finite.   

\newpage
\section{Additional results details}

\subsection{Proof of global optima in relaxed, resource-abundant, homogeneous scenario}
The below proof has been adapted from previous work \cite{Ramos-Fernandez2025}, being presented here with notation consistent with that of this paper.

Suppose that the population is homogeneous in foraging ability. Meaning, $N_k=N>0$ for $k\in [\mathcal{N}]$. Furthermore, assume that that $F\geq \sum_{i=1}^nN_i=nN$, so that the resource availability constraint (Equation \eqref{const: resources}) is trivial. Therefore for brevity we drop the final row of the constraint matrix $G$ and vector $\boldsymbol{h}$. In this section we prove that 
\begin{equation*}
    \boldsymbol{O}^* = \left(0, \dots, 0, \frac{N}{2}\right)
\end{equation*}
is the global maximiser of $T$ under the constraints $\boldsymbol{O}\geq 0$ and $G\boldsymbol{O}\leq \boldsymbol{h}$. We first simplify the coefficients of $T$, as derived in Section \ref{sec: coefficients}, for this scenario with homogeneous knowledge between individuals of the group. We then derive a condition for global maximality for our specific problem, which allows us to study `feasible perturbations' about a proposed optima. By analysing all possible kinds of feasible perturbation, we show that $\boldsymbol{O}^*$ is indeed a global maximiser. Our proof considers the relaxed scenario where problem variables and parameters are not constrained to be integers. We do arbitrarily suppose that $N$ is even, so that $N/2$ is an integer.
 
\begin{proposition}
    Suppose $N_k=N>0$ $\forall k\in [\mathcal{N}]$. Let $\hat{f}\colon [\mathcal{N}]\times [\mathcal{N}]\to \mathbb{N}$ be defined by 
    \begin{equation*}
        \hat{f}(i,j) =
            |I^{-1}(i)\cap I^{-1}(j)|
    \end{equation*}
    Then all of the coefficients in the objective function $T(\mathbf{O})$ can be expressed as
    \begin{align*}
        & l_i =N \sum_{k=2}^{f(i)}\binom{f(i)}{k} kN^{-k}, \\
        & M_{ij} = m_{ij,1}+m_{ij, 2}, \text{where:} \\
        & \hspace{40pt} m_{ij, 1} = \begin{cases}
            -2\sum_{k=2}^{\hat{f} (i, j)}\binom{\hat{f}(i, j)}{k}kN^{-k}\text{, if }\hat{f}(i,j) \geq 2   \\ 0, \text{ otherwise} 
        \end{cases} , \\ 
        & \hspace{40pt} m_{ij, 2} = \begin{cases}
            -\sum_{k=1}^{\hat{f}(i,j)}\binom{\hat{f}(i,j)}{k}(f(j)+k-1)N^{-f(i)-k} \text{, if } I^{-1}(j) \subset I^{-1}(i) \\ 
            0\text{, otherwise}
        \end{cases}\\
        &M_{ji} = M_{ij}
    \end{align*}
    where $i,j\in [\mathcal{N}]$ and $i\geq j$. 
    \begin{proof}
        First we deal with the linear terms given in the vector $\boldsymbol{l}=(l_i)$. In generality, these are given by
        \begin{equation*}
            l_i= \sum_{c\in J(i)} \left[\frac{\sum_{k\in I^{-1}(c)}N_k}{\prod_{k\in I^{-1}(c)}N_k}\right]
        \end{equation*}
        which under the homogeneity assumption (that $N_k=N$ $\forall c\in [\mathcal{N}]$), this simplifies to: 
        \begin{align}
            l_i= \sum_{c\in J(i)} \left[\frac{f(c)N}{N^{f(c)}}\right] = \sum_{c\in J(i)} f(c){N^{1-f(c)}}. 
        \end{align}
        The set $J(i)$ is the collection of indices $c\in[\mathcal{N}]$ such that the set $I^{-1}(c)\subseteq I^{-1}(i)$. Since $f(i)$ is the size of the set $I^{-1}(i)$, there are $\binom{f(i)}{k}$ many subsets of $I^{-1}(i)$ of size $k$ for $k=2,\dots, f(i)$. Since $l_i$ depends only upon the size of the subsets under the homogeneity assumption, we can group terms in the sum by their size. Doing this, we obtain the required expression
        \begin{equation*}
            l_i = \sum_{k=2}^{f(i)}\binom{f(i)}{k}kN^{1-k} = N\sum_{k=2}^{f(i)}\binom{f(i)}{ k}kN^{-k}.
        \end{equation*}

        We now compute $m_{ij, 1}$. This has the form:
        \begin{equation*}
            m_{ij, 1} = -\sum_{c\in V(i,j)}\frac{2f(c)}{\prod_{k\in I^{-1}(c)}N_k} = -2\sum_{c\in V(i,j)}f(c)N^{-f(c)}
        \end{equation*}
        The set $V(i,i)$, by definition, is given by 
        \begin{equation*}
            V(i,j) = \left\{c: I^{-1}(c)\subseteq I^{-1}(i)\cap I^{-1}(j) \right\}.
        \end{equation*}
        The size of the set $I^{-1}(i)\cap I^{-1}(j)$ is given by $\hat{f}(i,j)$ by definition. If $\hat{f}(i,j)=0$ or $\hat{f}(i,j)=1$, then $V(i,j)$ should be empty since we do not consider subgroups of size 0 or 1. If $\hat{f}(i,j)\geq 2$, then we can group terms in the sum by their corresponding subgroup size again, as done with the linear coefficients:
        \begin{equation*}
            m_{ij, 1}=  - 2\sum_{k=2}^{\hat{f}(i,j)}\binom{\hat{f}(i,j)}{k}kN^{-k}.
        \end{equation*}

        Now we focus on $m_{ij,2}$. Since $i\geq j$, we cannot have that $I^{-1}(i)\subset I^{-1}(j)$ by the definition of the index map $I$. If $I^{-1}(j)\not\subset I^{-1}(i)$ then $m_{ij,2}=0$. If $I^{-1}(j)\subset I^{-1}(i)$, then
        \begin{equation*}
            m_{ij,2} = -\sum_{c\in Z(i: j)}\frac{g(j)}{\prod_{k\in I^{-1}(c)}N_k} = -\sum_{c\in Z(i: j)}g(j)N^{-f(c)}.
        \end{equation*}
        The set $Z(i:j)$, by definition, is given by
        \begin{equation*}
            Z(i:j) = \{c:I^{-1}(j) \subset I^{-1}(c) \subseteq I^{-1}(i) \}
        \end{equation*}
        and therefore gives the number of `intermediate' sets between $I^{-1}(j)$ and $I^{-1}(i)$, including $I^{-1}(i)$ but excluding $I^{-1}(j)$. We therefore can break subsets again into their size, by noting that the number of intermediate sets of size will contain $f(j)$ and some additional elements from $f(i)$ not already contained in $f(j)$. The number of sets with $k$ additional elements will be $\binom{f(i)-f(j)}{k}$ for $k=1, \dots, f(i)-f(j)$. These sets will be of size $f(j)+k$. Therefore, in the case of $I^{-1}(j)\subset I^{-1}(i)$, we can write:
        \begin{equation*}
            m_{ij,2} = -\sum_{k=1}^{f(i)-f(j)}\binom{
            f(i)-f(j)}{k}(f(j)+k-1)N^{-f(j)-k}.
        \end{equation*}
        Then observing that $\hat{f}(i,j) = f(i) - f(j)$ under the condition $f(i)-f(j)$, we can express
        \begin{equation*}
            m_{ij,2} = -\sum_{k=1}^{\hat{f}(i,j)}\binom{\hat{f}(i,j)}{k}(f(j)+k-1)N^{-f(j)-k}.
        \end{equation*}
        Then $M_{ij} = m_{ij,1}+m_{ij,2}$ for $i\geq j$, giving all of the lower-diagonal entries. By the symmetry of $M$ we have all coefficients of $T$ simplified using the homogeneity assumption.

    \end{proof}
\end{proposition}

\begin{definition}
    The \textit{feasible region} is the set
    \begin{equation*}
        \Omega = \{\boldsymbol{O}\in \mathbb{R}^\mathcal{N} : G\boldsymbol{O}\leq \boldsymbol{h}, \boldsymbol{O} \geq 0\}.
    \end{equation*}
    Meaning, the set of points in which the constraints of the problem (as given in the previous section) are not violated. 
\end{definition}

\begin{definition}
    The \textit{feasible perturbation region about a point} $\boldsymbol{O}\in\mathbb{R}^\mathcal{N}$ is the set \begin{equation*}
        \Omega_\varepsilon(\boldsymbol{O}) = \{\boldsymbol{\varepsilon}\in\mathbb{R}^\mathcal{N}: \boldsymbol{O}+\boldsymbol{\varepsilon}\in \Omega\} 
    \end{equation*}
    Meaning, the set of perturbations from the point $\boldsymbol{O}$ which remain in the feasible region.  
\end{definition}

\begin{lemma}\label{lemma: optim cond}
    A feasible point $O^*\in\Omega$ is a local maximiser of $T$ if and only if there exists a neighbourhood $U\in \Omega_{\varepsilon}({O^*})$ around $\mathbf{0}$ such that every $ \varepsilon \in U$ satisfies
    \begin{equation} \label{eq: optimisation condition}
        T(\varepsilon) \leq -O^*M\varepsilon.
    \end{equation}
    Furthermore, if this property holds for $U= \Omega_{\varepsilon}({O^*})$, then $O^*$ is a global maximiser of $T$. We refer to this inequality as the optimality condition for $O^*$ with perturbation $\varepsilon$.

    \begin{proof}
        The condition for a feasible point $O^*\in\Omega$ to be a local maximiser of $T$ (\textit{the optimality condition}) is that there exists some $\epsilon>0$ such that $T(O) \leq  T(O^*)$ for all $O\in \Omega$ satisfying $\|O-O^*\|\leq \epsilon$.
        
        Any $O\in \Omega$ can be written as $O = O^* + \varepsilon$ for some $\varepsilon\in \Omega_\varepsilon(O^*)$, since both $O$ and $O^*$ are in $\Omega$ and by the definition of $\Omega_\varepsilon(O^*)$. 
        Using this expression for $O$, we can write $\|O-O^*\| = \|\varepsilon\|$. Therefore, the optimality condition can be rewritten as follows. A feasible point $O^*\in\Omega$ is a local maximiser of $T$ if there exists some $\epsilon>0$ such that $T(O^*+\varepsilon) \leq  T(O^*)$ for all $\varepsilon\in \Omega_\varepsilon(O^*)$ satisfying $\|\varepsilon\|\leq \epsilon$. This is equivalent to the point $O^*$ being optimal if there exists some neighbourhood $U\in \Omega_{\varepsilon}({O^*})$ around $\mathbf{0}$ such that every $ \varepsilon \in U$ satisfies
        $T(O^*+\varepsilon) \leq  T(O^*)$. Then observe that 
        \begin{align*}
            T(O^*+\varepsilon) &= \frac{1}{2}(O^{*t}+\varepsilon^t) M (O^{*}+\varepsilon) + l^t( O^{*}+\varepsilon) \\ 
            &=\frac{1}{2}O^{*t}MO^* + \frac{1}{2}O^{*t}M\varepsilon+\frac{1}{2}\varepsilon^tMO^* + \frac{1}{2}\varepsilon^tM\varepsilon + l^tO^{*} + l^t\varepsilon \\
            &=T(O^*) + T(\varepsilon) + \frac{1}{2}O^{*t}M\varepsilon+\frac{1}{2}\varepsilon^tMO^* \\ 
            &= T(O^*) + T(\varepsilon)+O^{*t}M\varepsilon, \text{ since } M \text{ is symmetric, so } O^{*t}M\varepsilon=\varepsilon^tMO^*.
        \end{align*}
        This implies the equivalence
        \begin{align*}
            T(O^* + \varepsilon) \leq T(O^*) & \iff T(O^* + \varepsilon)- T(O^*) \leq 0 \\ 
            & \iff T(\varepsilon)+O^{*t}M\varepsilon \leq 0 \\ & \iff T(\varepsilon) \leq -O^{*t}M\varepsilon. 
        \end{align*}
        Therefore, the optimality condition is equivalent to the statement that there exists some neighbourhood $U\in \Omega_{\varepsilon}({O^*})$ around $\mathbf{0}$ such that every $ \varepsilon \in U$ satisfies: 
        \begin{align}
            T(\varepsilon) \leq -O^{*t}M\varepsilon. 
        \end{align}
        Replacing $U$ with $\Omega_\varepsilon(O^*)$ throughout analogously gives the condition for global optimality. 
    \end{proof}
\end{lemma}

\begin{lemma}\label{lemma: first perturb}
Let $O^*=(0, \dots, 0, \frac{N}{2})$ and $\varepsilon=(\varepsilon_1,\dots, \varepsilon_\mathcal{N})\in \Omega_\varepsilon (O^*)$ be such that $\varepsilon_i\geq 0$  $\forall i \in [\mathcal{N}]$. Then $\varepsilon$ satisfies 
\begin{equation}\label{eq: optimality}
        T(\varepsilon) \leq -O^{*t}M\varepsilon
\end{equation}
  
\begin{proof}
    Let $\varepsilon=(\varepsilon_i)\in\Omega_\varepsilon (O^*)$ be any perturbation of the allowed form, where $\varepsilon_i \geq 0$ for $i=1,\dots, \mathcal{N}$. Define $v=(v_i) = O^{*t}M$. Then observe
    \begin{align*}
        v_i &= \sum_{j=1}^\mathcal{N} M_{ji}O_j \\
        &= M_{\mathcal{N}i}\frac{N}{2} \\
        &=\left(-2\sum_{k=2}^{f(i)}\binom{f(i)}{k}kN^{-k} -\sum_{k=1}^{n-f(i)}\binom{n-f(i)}{k}(f(i)+k-1)N^{-f(i)-k}\right) \frac{N}{2} \\
        & = -N\sum_{k=2}^{f(i)}\binom{f(i)}{ k}kN^{-k} -\frac{N}{2}\sum_{k=1}^{n-f(i)}\binom{n-f(i)}{k}(f(i)+k-1)N^{-f(i)-k} \\ & = -l_i - \frac{N}{2}\sum_{k=1}^{n-f(i)}\binom{n-f(i)}{k}(f(i)+k-1)N^{-f(i)-k} .
    \end{align*}
    Therefore, $v_i\leq -l_i$, since the second term in the right hand side of the final expression is strictly negative. This implies that 
    \begin{equation*}
        \sum_{i=1}^\mathcal{N} l_i\varepsilon_i \leq -\sum_{i=1}^\mathcal{N}{v_i}\varepsilon_i.
    \end{equation*}
    Therefore, since all of the elements of $M$ are non-positive, we also have that
    \begin{equation*}
        \frac{1}{2}\sum_{i=1}^\mathcal{N}\sum_{j=1}^\mathcal{N}\varepsilon_i\varepsilon_jM_{ij} + \sum_{i=1}^\mathcal{N} l_i\varepsilon_i \leq -\sum_{i=1}^\mathcal{N}{v_i}\varepsilon_i.
    \end{equation*}
    Which is equivalent to
    \begin{equation*}
        T(\varepsilon) \leq -v^t\varepsilon=-O^{*t}M\varepsilon
    \end{equation*}
    as required.
    
\end{proof}
\end{lemma}

\begin{lemma}\label{lemma: second perturb}
Let $O^*=(0, \dots, 0, \frac{N}{2})$ and $\varepsilon=(0,\dots, 0, -\varepsilon_\mathcal{N})$ be such that $\varepsilon_\mathcal{N}>0$. Then $\varepsilon$ satisfies $T(\varepsilon) \leq -O^{*t}M\varepsilon$.  
\begin{proof}
    Let $\varepsilon=(\varepsilon_i)\in\Omega_\varepsilon (O^*)$ be any perturbation of the allowed form, where $\varepsilon_i = 0$ for $i=1,\dots, \mathcal{N}-1$ and $\varepsilon_\mathcal{N} < 0$. Observe the following equivalence: 
    \begin{align*}
        T(\varepsilon) \leq -O^{*t}M\varepsilon & \iff \frac{1}{2}\varepsilon^{t}M\varepsilon+l^t\varepsilon \leq -O^{*t}M\varepsilon \\ 
        &\iff \sum_{i=1}^\mathcal{N}\sum_{j=1}^\mathcal{N} \varepsilon_iM_{ij}\varepsilon_j + \sum_{i=1}^\mathcal{N}\varepsilon_il_i \leq -\sum_{i=1}^\mathcal{N}\sum_{j=1}^\mathcal{N}O^*_iM_{ij}\varepsilon_j 
        \\& \iff \varepsilon_\mathcal{N}^2M_{\mathcal{N}\mathcal{N}}+ \varepsilon_\mathcal{N}l_\mathcal{N} \leq -O^*_\mathcal{N} M_{\mathcal{N}\mathcal{N}}\varepsilon_\mathcal{N} 
        \\& \iff \varepsilon_\mathcal{N}M_{\mathcal{N}\mathcal{N}}+ l_\mathcal{N} \geq -O^*_\mathcal{N} M_{\mathcal{N}\mathcal{N}} 
        \\&
        \iff \left(\varepsilon_\mathcal{N} +O^*\right)M_{\mathcal{N}\mathcal{N}} + l_\mathcal{N} \geq 0 
        \\&
        \iff -2\left(\varepsilon_\mathcal{N}+\frac{N}{2}\right) \left(\sum_{k=2}^n \binom{n}{ k}kN^{-k} \right)+ N\left(\sum_{k=2}^n\binom{n}{k}kN^{-k}\right) \geq 0 \\ 
        &\iff -2\varepsilon_\mathcal{N} -  N +  N\geq 0 \\
        & \iff \varepsilon_\mathcal{N} \leq 0
    \end{align*}
    which is true by assumption.
\end{proof}
\end{lemma}

\begin{remark}
    Any $\varepsilon \in \Omega_{O^*}$ can be written in the form of $\varepsilon^{(1)}$, as defined in Lemma \ref{lemma: first perturb}, or in the form of $\varepsilon^{(1)} + \varepsilon^{(2)}$, where $\varepsilon^{(2)}$ is as defined in Lemma \ref{lemma: second perturb}. These two cases correspond to the distinction between feasible perturbations in the positive and negative directions with respect to the final coordinate $O_\mathcal{N}$, respectively. Other coordinates must be perturbed from $O^*$ in the positive direction, since $O^*$ lies on the boundary of $\Omega$ (as $O_i = 0$ for $i=1,\dots, \mathcal{N}-1$). 
\end{remark}

\begin{theorem}\label{thm: optim}
    Let $O^*=\left(0, \dots, 0, \frac{N}{2}\right)$. Then $O^*$ is the global maximiser of $T$. 
\begin{proof}
    We construct a neighbourhood $U$ about $\boldsymbol{0}$ in $\Omega_\varepsilon(O^*)$ such that all $\varepsilon\in U$ satisfy Equation \eqref{eq: optimality}, which would imply that $O^*$ satisfies the optimality condition of Lemma \ref{lemma: optim cond} in the local sense. We then show that this constructed set must be equal to $\Omega_\varepsilon(O^*)$ itself, so that $O^*$ also satisfies the optimality condition of Lemma \ref{lemma: optim cond} in the global sense. To construct this set, we consider which perturbations in $\Omega_\varepsilon(O^*)$ satisfy inequality \eqref{eq: optimality}. As noted in the previous remark, each feasible perturbation can be written in the form of $\varepsilon^{(1)}$, or in the form of $\varepsilon^{(1)} + \varepsilon^{(2)}$. We consider these two cases separately.

    In the first case, all perturbations are of the form $\varepsilon = (\varepsilon_i)$ and $\varepsilon_i \geq 0$. By Lemma \ref{lemma: first perturb}, the condition \eqref{eq: optimality} holds for any such $\varepsilon\in\Omega_\varepsilon(O^*)$.

    In the second case, perturbations are of the form $\varepsilon = \varepsilon^{(1)} + \varepsilon^{(2)}$. Denote $\varepsilon^{(1)} = (\varepsilon^{(1)}_i)$ and $\varepsilon^{(2)} = (\varepsilon^{(2)}_i)$. Perturbations of this form will have $\varepsilon^{(1)}_i \geq 0$ for $i=1,\dots, \mathcal{N}$, $\varepsilon^{(2)}_j\geq 0$ for $i=1,\dots, \mathcal{N}-1$ and $\varepsilon^{(2)}_\mathcal{N}< 0$. Without loss of generality, assume that $\varepsilon^{(1)}_\mathcal{N} = 0$. Now observe that:
    \begin{align*}
        T(\varepsilon) &=   T\left(\varepsilon^{(1)}+ \varepsilon^{(2)}\right) \\
        & = \frac{1}{2}\left(\varepsilon^{(1)} + \varepsilon^{(2)}\right)^t M \left(\varepsilon^{(1)} + \varepsilon^{(2)}\right) + l^t\left(\varepsilon^{(1)} + \varepsilon^{(2)}\right) \\
        &=T\left(\varepsilon^{(1)}\right) + T\left(\varepsilon^{(2)}\right) + \varepsilon^{(1)t}M\varepsilon^{(2)} \\ 
        & \leq T(\varepsilon^{(1)})-O^{*t}M\varepsilon^{(2)} + \varepsilon^{(1)t}M\varepsilon^{(2)}\text{, by Lemma \ref{lemma: second perturb}} .
    \end{align*}
    Furthermore, by Lemma \ref{lemma: first perturb}, we have that
    \begin{equation*}
        T\left(\varepsilon^{(1)}\right) \leq -O^*M\varepsilon^{(1)},
    \end{equation*}
    which implies that 
    \begin{equation*}
        X \vcentcolon= -T\left(\varepsilon^{(1)}\right)-O^*M\varepsilon^{(1)}
    \end{equation*}
    is non-negative. Our current inequality for $T(\varepsilon)$ can be expressed as
    \begin{align*}
        T(\varepsilon) \leq -O^{*t}M\varepsilon^{(1)}-X-O^{*t}M\varepsilon^{(2)} + \varepsilon^{(1)t}M\varepsilon^{(2)}, 
    \end{align*}
    which implies that
    \begin{equation*}
        T(\varepsilon) \leq  -O^{*t}M\varepsilon +\left(-X +\varepsilon^{(1)t}M\varepsilon^{(2)}\right).
    \end{equation*}
    Therefore, $\varepsilon$ will satisfy the optimality condition \eqref{eq: optimality} if 
    \begin{equation*}
        -X +\varepsilon^{(1)t}M\varepsilon^{(2)} \leq 0 \iff X \geq \varepsilon^{(1)t}M\varepsilon^{(2)}.
    \end{equation*}
    Expanding the $\varepsilon^{(1)t}M\varepsilon^{(2)}$ term gives:
    \begin{equation*}
        \varepsilon^{(1)t}M\varepsilon^{(2)} = \sum_{i=1}^\mathcal{N}\sum_{j=1}^\mathcal{N}\varepsilon_j^{(1)}M_{ij}\varepsilon^{(2)}_i = \sum_{j=1}^\mathcal{N} \varepsilon_j^{(1)}M_{\mathcal{N}j}\varepsilon_\mathcal{N} = \varepsilon_\mathcal{N}\sum_{j=1}^\mathcal{N} \varepsilon_j^{(1)}M_{\mathcal{N}j}.
    \end{equation*}
    Therefore, $\varepsilon$ will satisfy the optimality condition \eqref{eq: optimality} if 
    \begin{equation*}
        X \geq  \varepsilon_\mathcal{N}\sum_{j=1}^\mathcal{N} \varepsilon_j^{(1)}M_{\mathcal{N}j} \implies \varepsilon_\mathcal{N} \geq \frac{X}{\sum_{j=1}^\mathcal{N} \varepsilon_j^{(1)}M_{\mathcal{N}j}},
    \end{equation*}
    using the fact that all entries of $M$ are negative and all elements of $\varepsilon^{(1)}$ are positive, which implies the above sum is negative. We use this fact to construct the set $U$.
    
    Consider the map $P\colon \Omega_\varepsilon(O^*) \to \mathbb{R}^\mathcal{N}$ which maps the final coordinate to zero and fixes the remaining coordinates. Define $U'= P(\Omega_\varepsilon(O^*))$ and
    \begin{equation*}
        K \vcentcolon= \min_{\varepsilon\in U'}\left[\frac{X}{\sum_{j=1}^\mathcal{N} \varepsilon_j^{(1)}M_{\mathcal{N}j}} \right],
    \end{equation*}
    which must exist, be finite and be non-zero since $N\neq 0$. Using the definition of $X$ we can write
    \begin{align*}
        K &= \min_{\varepsilon\in U'}\left[\frac{-T\left(\varepsilon^{(1)}\right)-O^*M\varepsilon^{(1)}}{\sum_{j=1}^\mathcal{N} \varepsilon_j^{(1)}M_{\mathcal{N}j}} \right] \\
        &  = \min_{\varepsilon\in U'}\left[\frac{-T\left(\varepsilon^{(1)}\right)-\frac{N}{2}\sum_{j=1}^\mathcal{N}\varepsilon_j^{(1)}M_{\mathcal{N}j}}{\sum_{j=1}^\mathcal{N} \varepsilon_j^{(1)}M_{\mathcal{N}j}} \right] \\ 
        & = \min_{\varepsilon\in U'}\left[\frac{-T\left(\varepsilon^{(1)}\right)}{\sum_{j=1}^\mathcal{N} \varepsilon_j^{(1)}M_{\mathcal{N}j}} \right] - \frac{N}{2}
    \end{align*}
    Now observe that the quantity in the minimisation operator can take the value 0, exactly when $T(\varepsilon) = 0$. We can construct a value of $\varepsilon\in U'\setminus \{0\}$ which solves $T(\varepsilon)$ as follows. Split $[n]$ into two \textit{disjoint} subsets, $S_1$ and $S_2$, consisting of $\left\lceil \frac{n}{2} \right\rceil$ and $\left\lfloor \frac{n}{2} \right\rfloor$ many individuals, respectively. Set $\varepsilon_j = N$ for $j = I(S_1)$ and $j = I(S_2)$, and all other coordinate values $\varepsilon_i=0$. Therefore, for all $c\in \left[\mathcal{N}\right]$, we have that either $P(c)=0$ (for the zero overlaps) or $A(c)=0$ (for the two non-zero overlaps of individuals, where the amount of unique knowledge is zero), which implies that the net information transfer, $T(\varepsilon)$, is $0$. Therefore, we can say 
    \begin{equation*}
        K \leq -\frac{N}{2}
    \end{equation*}
    Then, we can define $U_\mathcal{N}$ as
    \begin{equation*}
        U_\mathcal{N} = \{0\}\times \dots \times \{0\}\times \left[-\frac{N}{2}, \frac{N}{2}\right]
    \end{equation*}
    and then define $U$ as
    \begin{equation*}
        U =\left( U' + U_\mathcal{N}\right) \cap \Omega_{\varepsilon}(O^*)
    \end{equation*}
    where the $+$ denotes the Minkowski sum of sets (i.e the element-wise sum). By construction, $U$ is a neighbourhood in $\Omega_{\varepsilon}(O^*)$ about $\boldsymbol{0}$ where the condition 
    \begin{equation*}
        T(\varepsilon) \leq -O^{*t}M\varepsilon
    \end{equation*}
    holds for all $\varepsilon\in U$. Therefore, by Lemma \ref{lemma: optim cond}, $O^*$ is a local optimiser of $T$. For global optimality, it remains to show that $U = \Omega_{\varepsilon}(O^*)$. We do this by showing that 
    \begin{equation*}
        \Omega_{\varepsilon}(O^*) \subseteq U'+U_\mathcal{N}
    \end{equation*}
    which would imply that $U=\Omega_{\varepsilon}(O^*)$. Let $\varepsilon$ be any feasible perturbation. We show that this implies $\varepsilon\in U'+U_\mathcal{N}$. Write $\varepsilon = \varepsilon^{(1)}+\varepsilon^{(2)}$, where $\varepsilon^{(1)}$ contains the first $\mathcal{N}-1$ entries of $\varepsilon$ but has $\varepsilon^{(1)}_\mathcal{N}=0$, and $\varepsilon^{(2)}$ has $\varepsilon^{(2)}_i=0$ for all $i\in[\mathcal{N}]$ except for $\mathcal{N}$. We therefore note that, by definition, $\varepsilon^{(1)}=P(\varepsilon)$, so that $\varepsilon^{(1)}$ is in the image $ P(\Omega_\varepsilon (O^*))$. We show that $\varepsilon^{(2)}$ must be in $U_\mathcal{N}$. This is a straightforward argument. If $\varepsilon_\mathcal{N} < -\frac{N}{2}$, then $O^*+\varepsilon$ does not satisfy the constraint $\boldsymbol{O}\geq 0$: so no $\varepsilon\in\Omega_\varepsilon (O^*)$ can have this property. So, on the other hand, if $\varepsilon_{\mathcal{N}} > \frac{N}{2}$, then the condition $G(O^*+\varepsilon)\leq h$ can not be satisfied since
    \begin{align*}
    (O^*+\varepsilon)_\mathcal{N} = \frac{N}{2}+\varepsilon_\mathcal{N} > N,
    \end{align*}
    such that no $\varepsilon\in\Omega_\varepsilon (O^*)$ can have this property either. Hence, $\varepsilon^{(2)}\in U_\mathcal{N}$. Therefore, $\varepsilon\in U'+U_\mathcal{N}$. This implies that
    \begin{equation*}
        U = \left( U' + U_\mathcal{N}\right) \cap \Omega_{\varepsilon}(O^*) = \Omega_{\varepsilon}(O^*)
    \end{equation*}
    which therefore proves that $O^*$ is the global optimiser of $T$ over $\Omega$. 
\end{proof}
\end{theorem}

\subsection{Derivation of properties of optimally structured homogeneous systems}
In the previous subsection, we showed that the global solution to the optimisation problem (in Equations (2.3)-(2.4) of the main text) is given by: 
\begin{equation*}
    \boldsymbol{O}^* = \left(0,\dots, 0, \frac{N}{2}\right) \\
\end{equation*}
Here we compute the features ($w_n^{(i)}$, $K$ and $C_k$, as defined in the main text) of such optimal spatial structure. 

Firstly, we compute the relative overlap values corresponding to $\boldsymbol{O}^*$. Since the only size of subgroup with non-zero overlap is $n$, we have that $w_n^{(i)}=0$ for $i=2,\dots, n-1$. For $i=n$, corresponding to the overlap of the entire group, we have that
\begin{align*}
    w_n^{(n)}(\boldsymbol{O}^*) &= \frac{\sum_{c=1}^\mathcal{N}I_n(f(c))O_c}{ \sum_{i=1}^nN - \sum_{c=1}^\mathcal{N}(f(c)-1)O_c} \\ &
    = 
        \frac{O_\mathcal{N}}{nN -(n-1)O_\mathcal{N}} \\
        & = \frac{{N}/{2}}{nN-(n-1)N/2} \\
        & = \frac{1}{n+1}
\end{align*}
and therefore the proportion of points known by only one individual is given by:
\begin{equation*}
    w_n^{(1)} = 1-\sum_{i=2}^nw_n^{(i)} = 1-\frac{1}{n+1} = \frac{n}{n+1}.
\end{equation*}
Then we can compute the expected number of individuals with knowledge of each point directly as:
\begin{equation*}
    K(\boldsymbol{O}) = \sum_{i=1}^n i w_n^{(i)}(\boldsymbol{O}) = 1\cdot\frac{n}{n+1} + n\cdot \frac{1}{n+1} = \frac{2n}{n+1}.
\end{equation*}
Then we can compute the proportion of points that each individual shares with any other individual(s) as: 
\begin{equation*}
    C_k(\boldsymbol{O}) = \frac{\sum_{i\in \tilde{S}(k)}O_i}{N} = \frac{O_\mathcal{N}}{N} = \frac{N/2}{N} = \frac{1}{2}
\end{equation*}
for each $k\in [n]$. It remains to compute the value of the objective function at this point. Since only the final coordinate of $\boldsymbol{O}$ is non-zero, we have that:
\begin{align*}
    T(\boldsymbol{O}^*) &= \boldsymbol{l}^t\boldsymbol{O}+\frac{1}{2}\boldsymbol{O}^tM\boldsymbol{O} \\&= l_\mathcal{N}O_\mathcal{N}+\frac{1}{2}O_{\mathcal{N}}^2M_{\mathcal{NN}} \\
    &= \frac{N}{2}\left(N \sum_{k=2}^{f(\mathcal{N})}\binom{f(\mathcal{N})}{k} kN^{-k} \right) - \frac{1}{2}\left(\frac{N^2}{4}\right)\left(2\sum_{k=2}^{\hat{f} (\mathcal{N}, \mathcal{N})}\binom{\hat{f}(n, n)}{k}kN^{-k}\right) \\
    & = \frac{N^2}{2} \sum_{k=2}^{n}\binom{n}{k} kN^{-k} - \frac{N^2}{4} \sum_{k=2}^{n}\binom{n}{k} kN^{-k}  \\
    &= \frac{N^2}{4} \sum_{k=2}^{n}\binom{n}{k} kN^{-k} 
\end{align*}
where we have used the explicit forms of the coefficients of $\boldsymbol{l}$ and $M$ derived in the previous subsection. Observe that, when $N$ is large: 
\begin{align*}
    T(\boldsymbol{O}^*) &=  \frac{N^2}{4} \sum_{k=2}^{n}\binom{n}{k} kN^{-k} \\
    &= \frac{1}{2}\binom{n}{2} + \frac{N^2}{4} \sum_{k=3}^{n}\binom{n}{k} kN^{-k} \\
    &= \frac{n(n-1)}{4} + \sum_{i = 1}^{n-2} \binom{n}{i+2}(i+2)N^{-i} \\
   & = \frac{n(n-1)}{4} + \mathcal{O}(N^{-1})
\end{align*}
where in the third line we have re-indexed the sum with $i=k-2$. 

\clearpage
\subsection{Additional figures}

\begin{figure}[ht]
    \centering
    \includegraphics[width=\linewidth]{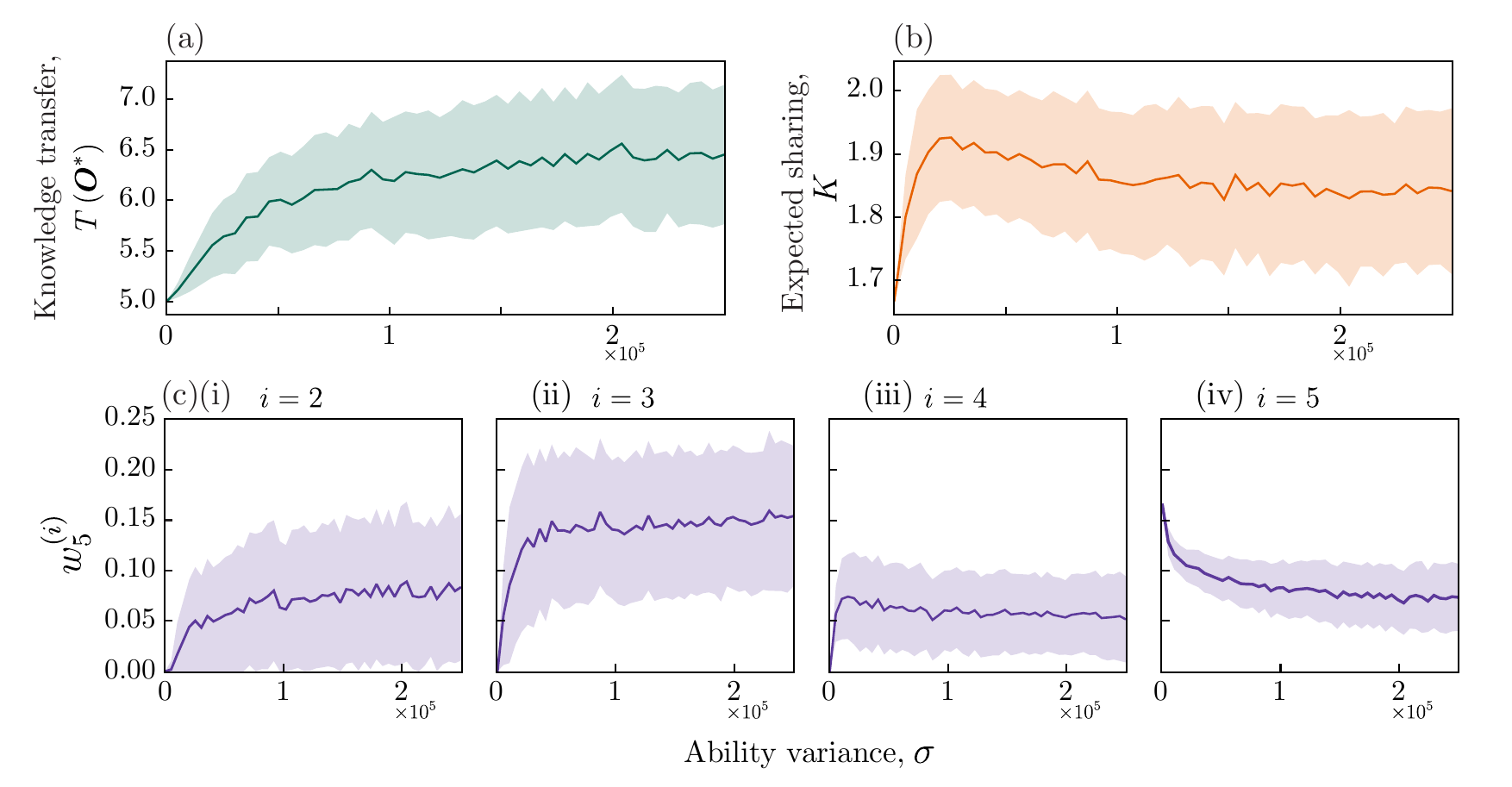}
    \caption{Results from our general heterogeneity scenario in the resource-abundant case with $\sigma\in [0, 2.5\times 10^5]$, mean foraging ability (of the distribution generating $\boldsymbol{N}$) $N=10^4$ and group size $n=5$. Shown are: (a) the value of the objective function at the optimal solution, $T(\boldsymbol{O}^*)$, (b) the expected number of individuals with knowledge of each point under the optimal structure, $K$, (c) the relative orders of spatial overlap, $w_5^{(i)}$, for (i) $i=2$, (ii) $i=3$, (iii) $i=4$ and (iv) $i=5$. In each plot, solid lines are the average value across each of 250 Monte Carlo simulations for each value of $\sigma$, and shaded regions show one standard deviation from this point.}
    \label{fig:7_wide}
\end{figure}

\begin{figure}[ht]
    \centering
    \includegraphics[width=\linewidth]{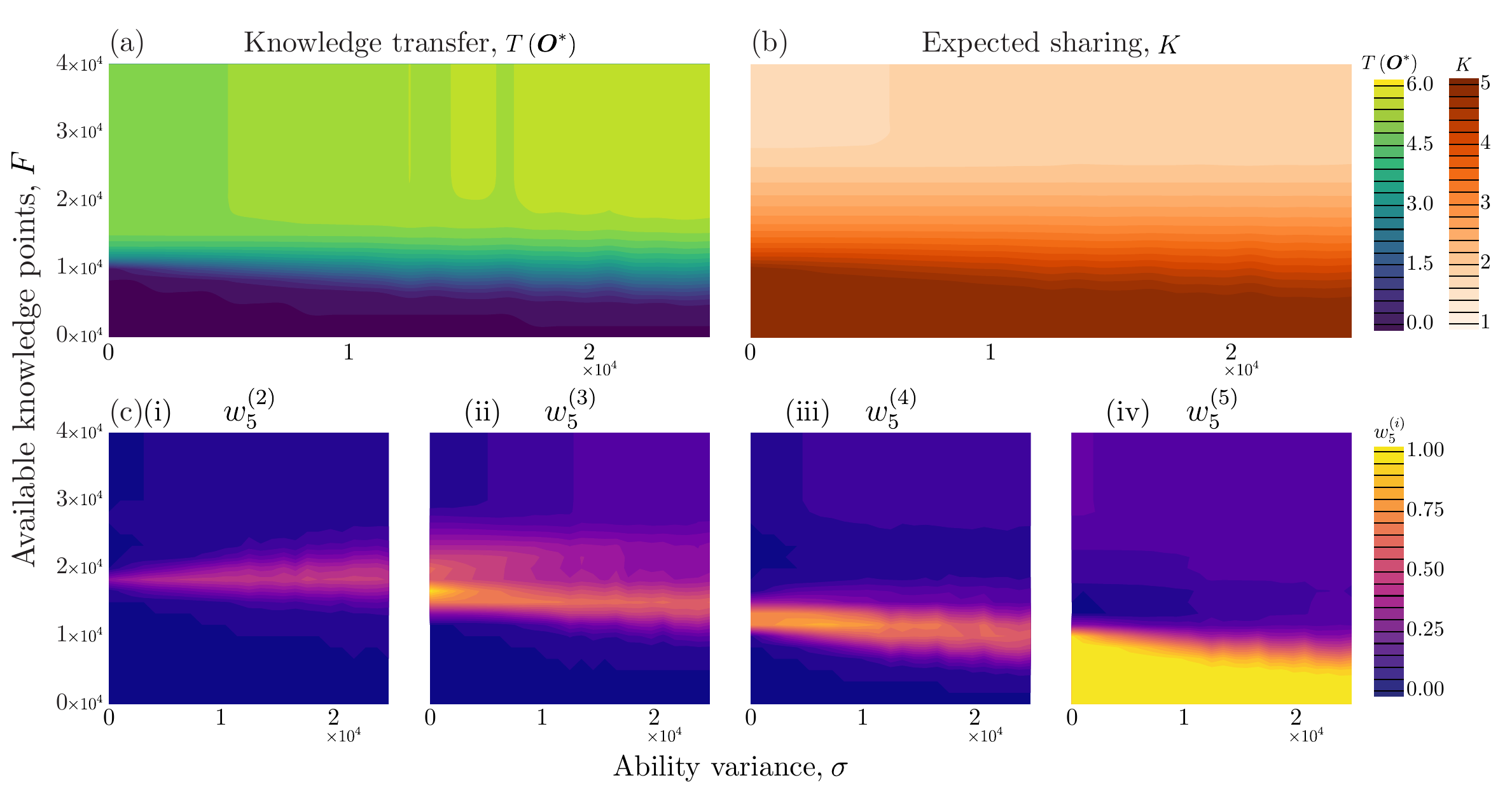}
    \caption{Results from our general heterogeneity scenario with $\sigma\in [0, 2.5\times 10^4]$, mean foraging ability $N=10^4$, foraging constraint varying between $F\in [0, 4\times 10^4]$ and for group size $n=5$. Specific values of $\sigma$ and $F$ are obtained from 25 uniformly spaced points over their respective intervals, each rounded upwards. Shown are: (a) the value of the objective function at the optimal solution, $T(\boldsymbol{O})$, (b) the expected number of individuals with knowledge of each point under the optimal structure, $K$, (c) the relative orders of spatial overlap $w_5^{(i)}$ for (i) $i=2$, (ii) $i=3$, (iii) $i=4$ and (iv) $i=5$. In each contour plot, every value is the average value across each of the 100 Monte Carlo simulations for the corresponding value of $\sigma$ and $F$. }
    \label{fig:8_norm}
\end{figure}

\begin{figure}[ht]
    \centering
    \includegraphics[width=\linewidth]{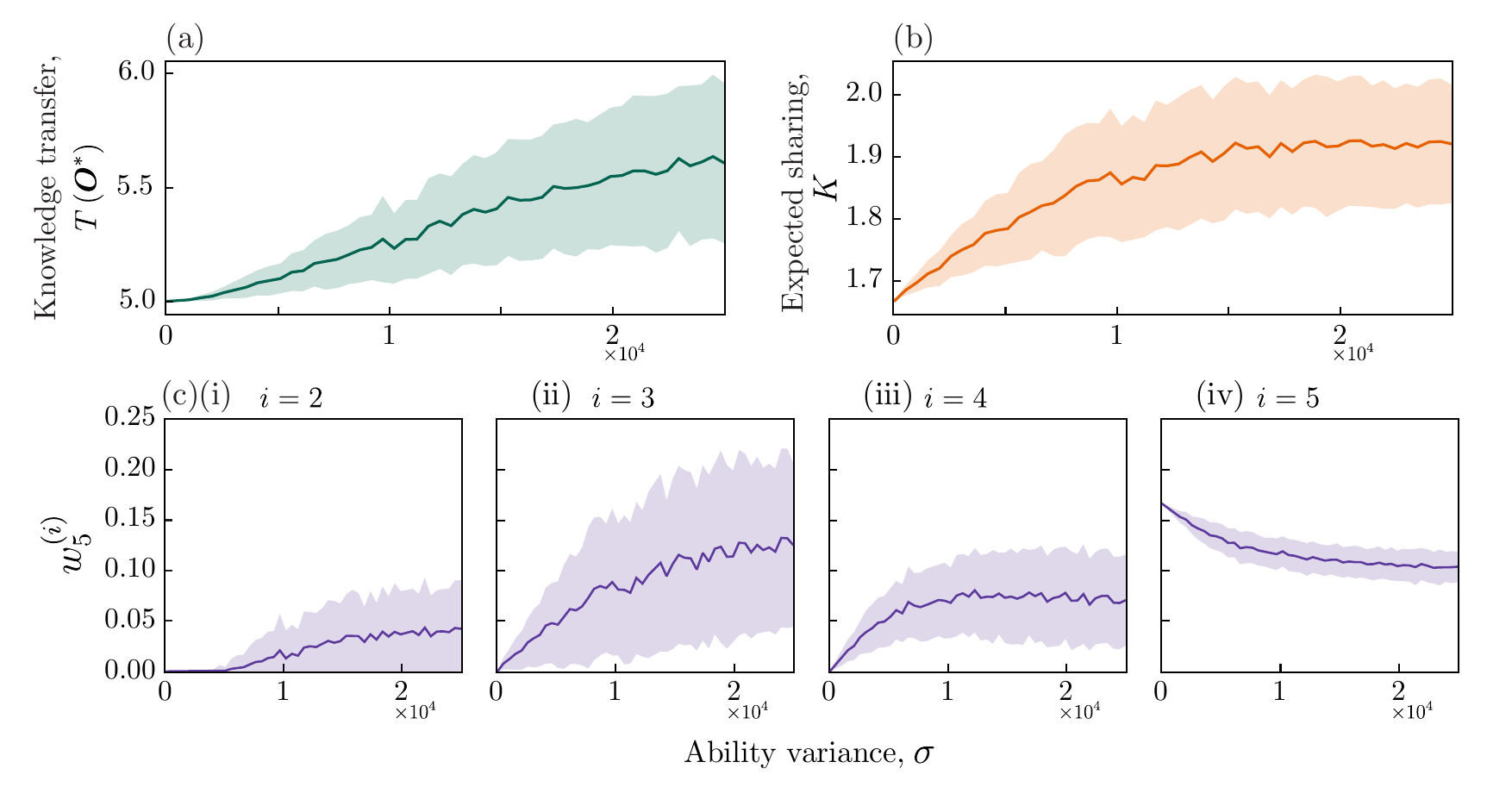}
    \caption{Results from our general heterogeneity scenario in the resource-abundant case with $\sigma\in [0, 2.5\times 10^4]$, mean foraging ability (of the distribution generating $\boldsymbol{N}$) $N=10^4$ and group size $n=5$. Each individual collection of group abilities, $\boldsymbol{N}$, have been normalised to have mean $N$. Shown are: (a) the value of the objective function at the optimal solution, $T(\boldsymbol{O}^*)$, (b) the expected number of individuals with knowledge of each point under the optimal structure, $K$, (c) the relative orders of spatial overlap, $w_5^{(i)}$, for (i) $i=2$, (ii) $i=3$, (iii) $i=4$ and (iv) $i=5$. In each plot, solid lines are the average value across each of 250 Monte Carlo simulations for each value of $\sigma$, and shaded regions show one standard deviation from this point.}
    \label{fig:7_norm}
\end{figure}

\clearpage
\printbibliography